\documentclass[lettersize,journal]{IEEEtran}
\usepackage{comment}
\usepackage{calc,amsfonts,amssymb,amsmath,bm,url,color,theorem,graphicx,cite}
\usepackage{psfrag,subfigure,float}
\usepackage{multirow,subfigure,amsmath,epsfig,amsfonts,amssymb,psfig,graphics,psfrag,theorem,calc,subfigure,url,bm,cite}
\usepackage{stfloats,hyperref}  
\usepackage[table]{xcolor}
\usepackage{array}
\usepackage{algorithm,algorithmic}
\usepackage{diagbox} 
\usepackage{hhline}
\usepackage{caption}
\usepackage{bbding}
\usepackage{amssymb}
\usepackage{booktabs}
\usepackage{textcomp}
\usepackage{threeparttable}
\usepackage{graphicx}
\usepackage{multirow}
\usepackage{makecell}
\usepackage{booktabs}
\usepackage{tikz}
\usepackage[table]{xcolor}
\usepackage{booktabs}
\usetikzlibrary{quantikz}
\usepackage{mathtools}

\def\blue{\color{blue}}

\usepackage{colortbl}

\makeatletter
\def\multilimits@{\bgroup
	\Let@
	\restore@math@cr
	\default@tag
	\baselineskip\fontdimen10 \scriptfont\tw@
	\advance\baselineskip\fontdimen12 \scriptfont\tw@
	\lineskip\thr@@\fontdimen8 \scriptfont\thr@@
	\lineskiplimit\lineskip
	\vbox\bgroup\ialign\bgroup\hfil$\m@th\scriptstyle{##}$\hfil\crcr}
\def\Sb{_\multilimits@}
\def\endSb{\crcr\egroup\egroup\egroup}
\makeatother

{\begin{list}{}%
		{\setlength{\rightmargin}{0pc}%
			\setlength{\leftmargin}{1pc} }} %
	{\end{list}}

\newlength{\twidth}
\definecolor{orange}{RGB}{255,107,0}
\def\blue{\color{blue}}

\newtheorem{Assumption}{Assumption}
\newtheorem{Lemma}{Lemma}

\theorembodyfont{\rmfamily}

\newtheorem{Remark}{Remark}

{\begin{list}{}{
			\settowidth{\labelwidth}{\mbox{\textnormal{#1}}}%
			\setlength{\leftmargin}{\labelwidth+\labelsep}}}%
	{\end{list}}

\newcommand\bA{\ensuremath{{\bm A}}}
\newcommand\bB{\ensuremath{{\bm B}}}

\newcommand\bD{\ensuremath{{\bm D}}}

\newcommand\bQ{\ensuremath{{\bm Q}}}

\newcommand\bS{\ensuremath{{\bm S}}}

\newcommand\bU{\ensuremath{{\bm U}}}

\newcommand\bX{\ensuremath{{\bm X}}}

\newcommand\bZ{\ensuremath{{\bm Z}}}

\newcommand\bd{\ensuremath{{\bm d}}}

\newcommand\bx{\ensuremath{{\bm x}}}
\newcommand\by{\ensuremath{{\bm y}}}
\newcommand\bz{\ensuremath{{\bm z}}}

\definecolor{orange}{RGB}{255,107,0}
\def\blue{\color{black}}

\begin{document}

\title{Anti-Hyperspectral Anomaly Detection: A First Study on Stealthy Lipschitz-Forcing Perturbations Against Unknown Detectors
}

\author{Chia-Hsiang Lin,~\IEEEmembership{Senior Member,~IEEE}, Si-Sheng Young,~\IEEEmembership{Student Member,~IEEE},
\\
and Jon Atli Benediktsson,~\IEEEmembership{Life Fellow,~IEEE}
\thanks{This study was partly supported by the Emerging Young Scholar Program (namely, the 2030 Cross-Generation Young Scholars Program) of National Science and Technology Council (NSTC), Taiwan, under Grant NSTC 114-2628-E-006-002.
   We thank the National Center for Theoretical Sciences (NCTS) and the National Center for High-performance Computing (NCHC) for providing the computing resources.}

    \thanks{\textit{(Co-corresponding authors: Chia-Hsiang Lin, and Jon Atli Benediktsson)}}
    \thanks{C.-H. Lin is with the Department of Electrical Engineering, National Cheng Kung University, Tainan, Taiwan (R.O.C.) (e-mail: chiahsiang.steven.lin@gmail.com).}
    \thanks{S.-S. Young is with the Institute of Computer and Communication Engineering, Department of Electrical Engineering, National Cheng Kung University, Tainan, Taiwan (R.O.C.) (e-mail: q38121509@gs.ncku.edu.tw).}
    \thanks{J. A. Benediktsson is with the Faculty of Electrical and Computer Engineering, University of Iceland, Reykjavík, Iceland (e-mail: benedikt@hi.is).}
 }

\maketitle

\begin{abstract}
Hyperspectral imagery represents the most advanced contemporary technology for remotely detecting anomalous objects.
Nevertheless, hyperspectral anomaly detection (HAD) technique makes ground facilities/situations completely exposed. 
For the first time, we develop the first anti-HAD (AHAD) technique rendering the key objects undetected, without perfect coordinate/position state information (CSI) of the detectors (e.g., reconnaissance aircraft).
Our AHAD algorithm is generally applicable for defending against mainstream HAD baselines, including data-driven and model-driven methods. 
AHAD is fundamentally different from conventional adversarial attacks, so a novel theory is required. 
We customize novel regularizers to assimilate real anomalies into the backgrounds (ARAB) and fool the detectors with pseudo-anomalies, thereby optimizing an energy-efficient, stealthy perturbation signal for AHAD. 
The ARAB regularization can be mathematically interpreted as flattening the topology-enhanced anomaly/background structures in the feature space, hence termed Lipschitz-forcing perturbations.
Considering the imperfect CSI, we further develop a robust AHAD criterion in which the uncertainty is mathematically described as matrix-shifting misalignment for statistically generating the robust perturbation. 
Comprehensive experiments demonstrate the effectiveness and robustness of our AHAD algorithm across diverse real-world datasets. 
{\blue Remarkably, our algorithm generates a single AHAD perturbation signal that can simultaneously evade over half of benchmark detectors}, demonstrating its strong potential for real-world AHAD applications, given that the reconnaissance detector type is usually unknown. 
To the best of our knowledge, this is the first formal AHAD study.
As a side contribution, we propose a new quantitative performance index, ArmCBA, to evaluate the robustness of an HAD method against our AHAD signal.
The demo code is available at \url{https://github.com/IHCLab/AHAD}. 
\end{abstract}

\begin{IEEEkeywords}
Hyperspectral anomaly detection, 
hyperspectral remote sensing, 
anti-reconnaissance method, 
anti-detection method,
adversarial attack.
\end{IEEEkeywords}

\section*{Nomenclature}
\begin{IEEEdescription}[
    \IEEEusemathlabelsep
    \IEEEsetlabelwidth{ArmCBA}
]
 \item[AHAD] Anti-hyperspectral anomaly detection.
    \item [ARAB] Assimilating real anomalies into the backgrounds.
     \item[ArmCBA] Area missing caused by anti-detection, defined as $(1-\frac{\text{AUC}_{ap}}{\text{AUC}_{up}})\times 100\%~~\!(\uparrow)$.
     \item[AVIRIS] Airborne Visible/Infrared Imaging Spectrometer.
      \item[AA] Adversarial attack.
    \item[AI] Artificial intelligence.
    \item[ASF] Affine-set fitting. 
      \item[AUC] Area under the curve.
     \item[$\mathrm{AUC}_{up}$] AUC score obtained from the unperturbed HSI.
    \item[$\mathrm{AUC}_{ap}$] AUC score obtained from the AHAD-perturbed HSI.
    \item[BSS] Blind source separation.
     \item[CSI] Coordinate/position state information.
     \item[CNN] Convolutional neural network.
    \item[DL] Deep learning.
      \item[D2C] Data-to-center distance.
     \item[D2CM] Data-to-center-weighted mask.
   \item[EE] Energy efficiency.
    \item[HAD] Hyperspectral anomaly detection.
     \item[HSI] Hyperspectral image.
     \item[HAA] Hyperspectral adversarial attack.
     \item[i.i.d.] Independent and identically distributed.
      \item[LRR] Low-rank representation.
     \item[MIMO] Multi-input multi-output.
     \item[NASA] National Aeronautics and Space Administration.
     \item[P2M] Point-to-manifold distance.
     \item[PDirs] Principal directions.
     \item[PF] Probability of false alarm.
    \item[PD] Probability of detection.
     \item[PAG] Pseudo-anomaly generation.
     \item[QUEEN] Quantum deep network.
     \item[RS] Remote sensing.
      \item[RPCA] Robust principal component analysis.
     \item[RR] Restoration resistance.
      \item[REG] Regularization. 
     \item[SNR] Signal-to-noise ratio.
    \item[SOTA] State-of-the-art.
    \item[SSTV] Spectral-spatial total variation.
    \item[SV] Singular value.
    \item[SL] Supervised learning.
     \item[TV] Total variation.
    %
    
\end{IEEEdescription}

\begin{figure}[t]
    \centering
    \includegraphics[width=1\linewidth]{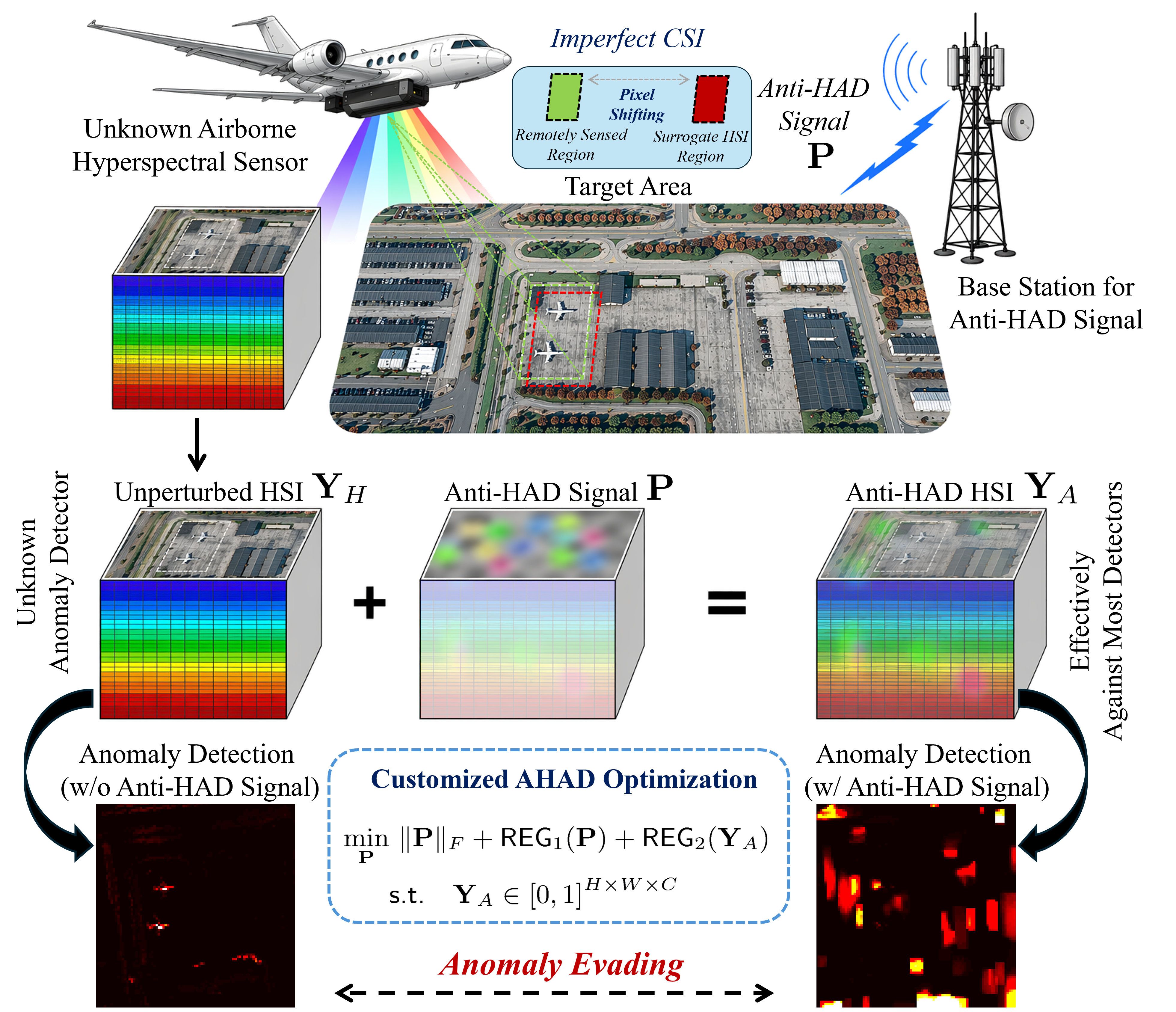}
    \caption{Schematic illustration of the anti-hyperspectral anomaly detection (AHAD) problem.
    Hyperspectral anomaly detection techniques enable the identification of anomalous objects under a fully blind setting. 
    However, when an unknown hyperspectral sensor observes a target area, such advanced detection capabilities expose our ground facilities and activities. 
    To address this security-critical limitation, this study formally defines the AHAD problem and, to the best of our knowledge, develops the first AHAD solution.
    Physical restrictions behind AHAD, such as imperfect coordinate/position state information (CSI), are introduced in Section \ref{sec: Intro}
    }\label{fig: AHAD_concept}
\end{figure}
\section{Introduction}\label{sec: Intro}
\begin{figure*}[t]
    \centering
    \includegraphics[width=1\linewidth]{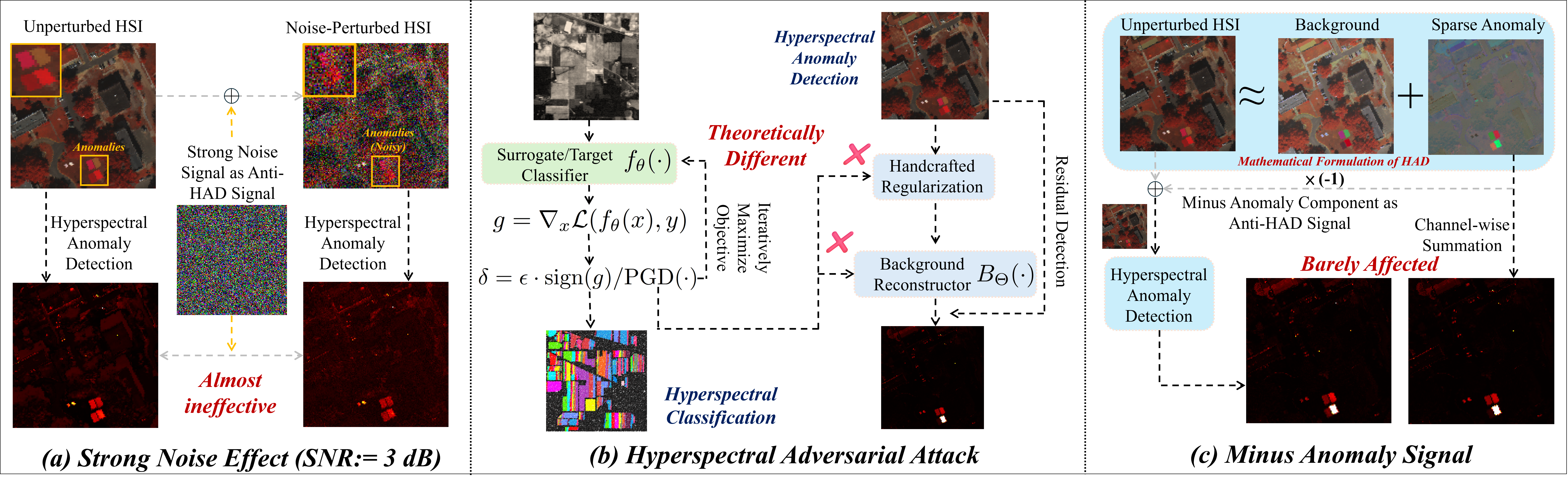}
    \caption{Schematic illustration showing why the existing perturbation/interference schemes, including (a) strong noise effect, (b) hyperspectral adversarial attack, and (c) minus anomaly signal, are inadequate for the challenging AHAD task, even when high-energy perturbation signals are applied.
    These limitations motivate the development of a radically new AHAD framework.
    More details can be found in Section \ref{sec: Intro} and Section \ref{sec: Related Work}.
    }\label{fig: intro_fig}
\end{figure*}
In modern remote sensing (RS), advanced hyperspectral imaging systems can acquire data over a broad spectral range \cite{HyFuHAD,hong2026hyperspectral}, and the resulting spectral signatures are unique across different materials, enabling advanced object identification \cite{8697135} and detection \cite{TGFA-AD}. 
Among diverse RS applications, hyperspectral anomaly detection (HAD), as a security-critical application \cite{SuperRPCA,6784134}, has been extensively studied for detecting distinct spectral signatures (potential threats) relative to their background under a blind setting.
For example, deep learning (DL)-based HAD algorithms \cite{Auto-AD,NL2Net,BockNet} have been widely developed ({see} Section \ref{subsec: HAD}).
Such advanced techniques fully expose ground facilities and situations, leaving no place to hide for suspicious anomalies.
Although HAD techniques ensure a high level of monitoring capability, the resulting advantages can be seen as a double-edged sword.

When unknown hyperspectral sensors (e.g., reconnaissance aircraft) pass over a target area on our side ({see} {Fig.}~\ref{fig: AHAD_concept}), the associated ground facilities and activities are just exposed under the opponent's HAD technique.
Unfortunately, this critical and risky RS \textit{dilemma} has not yet been discussed.
More worryingly, none of the existing techniques can be used effectively as remedial actions.
With the above strong motivations, this paper introduces, for the first time, the concept of the anti-HAD (AHAD), together with the first customized AHAD technique ({see} {Fig.}~\ref{fig: AHAD_concept}) to address this challenging problem.
More detailed problem definitions and practical challenges, such as imperfect coordinate/position state information (CSI) of sensors, are discussed in Section \ref{sec: Proposed}.

Briefly explaining, let $\mathbf{Y}_H\in\mathbb{R}^{H\times W\times C}$ denote a $C$-band HSI with $L=HW$ pixels of the target area, AHAD technique aims to estimate a perturbation signal $\mathbf{P}\in\mathbb{R}^{H\times W\times C}$ such that the resulting AHAD image $\mathbf{Y}_A=\mathbf{Y}_H+\mathbf{P}$ renders the key objects undetected ({see} {Fig.}~\ref{fig: AHAD_concept}).
Since we do not know the opponent's hyperspectral detectors, we need to compute a universal $\mathbf{P}$ that can interfere with almost all the mainstream HAD methods, thereby achieving reliable protection.
Although AHAD appears mathematically simple at first glance, it is in fact highly challenging and requires a fundamentally different theory, as discussed below.

First of all, the actual HSI sensed by the opposing sensors (e.g., reconnaissance aircraft) is typically inaccessible.
To estimate the perturbation signal, obtaining a surrogate HSI\footnote{In the black-box adversarial attack, the target model is inaccessible, and an approximate network, called a surrogate network \cite{ma2025efficient}, is used to estimate the perturbation. Similarly, as the target HSI is generally unavailable in the AHAD task, we refer to the available approximate HSI as a surrogate HSI.} $\widetilde{\mathbf{Y}}_H\in\mathbb{R}^{H\times W\times C}$ to approximate $\mathbf{Y}_H$ is essential.
Constructing $\widetilde{\mathbf{Y}}_H$ is relatively straightforward (e.g., using HSI synthesis \cite{10000408}) when the CSI of the opposing sensors is known.    
Specifically, comprehensive information on the ground facilities and activities within the anti-detecting area is accessible on our side.
Moreover, the hardware configurations of contemporary hyperspectral sensors are often available or can be reasonably estimated \cite{cocks1998hymaptm}; for instance, their spectral ranges may be reasonably assumed to be some subsets of existing high-standard hyperspectral systems (e.g., NASA's AVIRIS \cite{AVIRISdata}).
In this context, the perfect CSI (e.g., sensor positions, imaging geometry, and viewing angles) enables the construction of $\widetilde{\mathbf{Y}}_H$ via specific techniques (e.g., sensor pose estimation \cite{shetty2019uav}, viewing image synthesis \cite{ding2020practical,li2026unifying,vuong2025aerialmegadepth}), which serves as an alternative to $\mathbf{Y}_H$ and facilitates subsequent AHAD signal estimation.
More details are provided in Section \ref{subsec: perfect CSI Formulation}.

In practice, however, CSIs of the reconnaissance sensors must be estimated \cite{934067,toth2010airborne} and therefore remain imperfect.
This imperfection presents a challenge similar to those encountered in data fusion (i.e., pixel shifting), but remains significantly more challenging.
In detail, multimodal RS data often require additional pixel-level calibration for accurate data fusion \cite{7399718,QRCODE}.
However, the calibration strategy remains impractical for AHAD.
As a result, the imperfect CSI manifests as uncertainty in spatial alignment (i.e., pixel shifting), as illustrated in the green and red regions in {Fig.}~\ref{fig: AHAD_concept}.
Consequently, after constructing the $\widetilde{\mathbf{Y}}_H$, a practically applicable AHAD framework should produce $\mathbf{P}$ that is robust against spatial misregistration, unless the CSI can be perfectly known.

\subsection{Differences Between AHAD and Conventional Attacks}

After introducing the AHAD problem, we present some seemingly related problems (e.g., adversarial attacks) and discuss their fundamental differences, suggesting that AHAD is a new topic and requires a new solution.

For the design of the AHAD signal $\mathbf{P}$, a primary and fundamental requirement is to be generally applicable for defending against almost all mainstream HAD methods (uncertainty of the detectors), as aforementioned. 
Furthermore, we have the uncertainty caused by the imperfect CSI.
We emphasize that even with perfect CSI, constructing a reliable $\mathbf{P}$ remains highly non-trivial.

Subsequently, we explain why almost all existing interference/perturbation schemes ({see} {Fig.}~\ref{fig: intro_fig}), including strong noise effects, hyperspectral adversarial attacks (HAA), and the use of minus anomaly signals, are inapplicable to this AHAD task.
First, as shown in {Fig.}~\ref{fig: intro_fig}, an intuitive approach to AHAD is to treat $\mathbf{P}$ as a high-energy noise signal; however, this is impractical because existing HAD approaches are robust to a certain (or even relatively high) noise level.
To substantiate this robustness, we compare the detection results of a mainstream HAD method \cite{RGAE} obtained from an unperturbed HSI and its extremely noise-perturbed counterpart (SNR:= 3 dB).
As presented in {Fig.}~\ref{fig: intro_fig}(a), such strong noise perturbations have only a limited impact on the detection performance; even if the noise degrades performance, well-established hyperspectral noise-removal/restoration techniques \cite{HyperKING,HyperQUEEN,9318503} can be used to recover clean spectral signatures.   
These facts, taken together, hinder the noise-based AHAD strategy.

As for the HAA [{see} {Fig.}~\ref{fig: intro_fig}(b)] and minus anomaly signals [{see} {Fig.}~\ref{fig: intro_fig}(c)] schemes, the former is primarily developed to attack hyperspectral classification baselines, while the latter produces solely a small residual signal that has negligible effects on the HAD process.
Neither scheme can be used to effectively defend HAD methods.
Since these interference schemes require more background knowledge to understand their theoretical limitations, we refer readers to Section \ref{sec: Related Work} for a more detailed discussion.
Consequently, none of the existing approaches is suitable for accomplishing the AHAD tasks, necessitating the development of a radically new theory.

\subsection{Backgrounds of the Proposed AHAD Solution}

In this study, a solution for the AHAD task is to assimilate real anomalies into background components (ARAB).
To be specific, given an HSI, anomalies refer to pixels with distinct spectral characteristics relative to their surrounding background \cite{SuperRPCA,TGFA-AD}.
In light of this, anomalies and background pixels are expected to differ substantially in terms of Euclidean distance or gradient.
To accomplish ARAB, the objective can be mathematically interpreted as to flattening the topologies of the perturbed structures.
When the perturbed data topologies exhibit no clearly distinguishable anomalous pixels, we empirically found that mainstream detectors tend to identify all spectral signatures as background, a process referred to as ARAB.
Moreover, by flattening the topology-enhanced structures (rather than the original structures; {see} Section \ref{sec: Proposed}), we can more effectively assimilate the anomalies into the backgrounds, thereby facilitating the ARAB process.
However, in the original hyperspectral domain, the ARAB aims to minimize spectral variations across high-dimensional signatures, resulting in a substantial computational burden due to the hundreds of bands.

%

%
Fortunately, remotely sensed HSIs have been shown to lie in an $N$-dimensional subspace (with $N\ll C$) due to their low-rank nature.
An appropriate feature extraction or dimension reduction technique, such as subspace identification \cite{SISHY,HyperQUEEN}, can effectively reduce the spectral redundancy while preserving informative structure.
Therefore, we perform an efficient ARAB process in the feature space, rather than in the original hyperspectral domain. 
Moreover, in the feature space, we observe that anomalies are typically distributed far from the data mean, which motivates us to construct a data-to-center-weighted mask (D2CM).
The D2CM enhances the data topology by radially scaling each data point based on its data-to-center distance (D2C), thereby amplifying the variations of real anomalies and making ARAB more targeted (to be detailed in Section \ref{sec: Proposed}).
Building on these observations, we propose an ARAB regularization that minimizes the discrete gradient of the D2CM-topology-enhanced structures in the feature space.
Once anomalies exhibit only low disparity (relative to background) in the topology-enhanced structure, they naturally become indistinguishable from background components, even without D2CM.
Echoing the Lipschitz constraint in DL models, we term these AHAD signals as Lipschitz-forcing perturbations.  

On the other hand, certain advanced detectors may still identify anomalies, even when the difference between anomaly and background pixels is significantly reduced (e.g., after applying ARAB regularization).
To further fool these detectors, we additionally introduce the pseudo-anomaly effect.
In detail, real-world anomalies often correspond to a small group of substances that deviate from the dominant background components.
Therefore, amplifying the energy of the tail (non-principal) components allows us to generate noticeable pseudo-anomalies.
Taking the above ARAB regularization and pseudo-anomaly generation strategy together, the proposed AHAD framework is able to generate an AHAD perturbation signal that demonstrates considerable potential to evade most mainstream detectors under the perfect CSI assumption. 

Finally, as we mentioned above, the imperfect CSI, which manifests as an uncertainty in the pixel-shifting issue between $\widetilde{\mathbf{Y}}_H$ and $\mathbf{Y}_H$, should be considered to facilitate practical applicability. 
To address this limitation, we further develop a robust AHAD criterion based on stochastic optimization ({see} Lemma \ref{Lemma 1}). 
Specifically, when a reconnaissance sensor passes through the target area (our side), the corresponding $\widetilde{\mathbf{Y}}_H$ must be constructed based on an estimated CSI with the highest confidence/probability.
Under this assumption, a larger pixel shifting indicates a greater deviation between the estimated CSI and the true CSI of the unknown sensor.
In practice, these worst scenarios (i.e., severe pixel shifting) are expected to retain lower confidence/probability under the estimated CSI model, because the most confident CSI (which is expected to be a no-shift case) has been employed to construct $\widetilde{\mathbf{Y}}_H$.
Therefore, it is reasonable to assume that the spatial misalignment caused by imperfect CSI follows a normal distribution; more details are provided in Section \ref{subsec: Robust}.
With this prior distribution of CSI uncertainty, a robust perturbation signal (against pixel-shifting) can be naturally achieved by minimizing the expectation (with respect to all possible $\widetilde{\mathbf{Y}}_H$) of the AHAD criterion.
Then, inspired by the 5G/6G techniques (channel uncertainty, in particular) \cite{chi2017convex}, the induced probabilistic objective function is transformed into a deterministic one, in order to facilitate the subsequent optimization procedure ({see} Lemma \ref{Lemma 1}).
With the above customized designs, the proposed AHAD solution is hence advanced as a fully model-free, annotation-free, unsupervised framework.
These advantages enable us to realistically construct a single perturbation signal that exhibits wide transferability across mainstream detectors and is highly robust across numerous pixel-shifting cases. 
To the best of our knowledge, this article presents the first AHAD study.

The remainder of this article is structured as follows.
In Section \ref{sec: Related Work}, we review the limitations of the existing HAA approaches in addressing AHAD tasks, and introduce mainstream HAD algorithms.
Next, Section \ref{sec: Proposed} mathematically presents the problem definitions, formulations, and implementation details of the proposed AHAD framework.
Section \ref{sec: Experiment} demonstrates the effectiveness and robustness of our AHAD algorithm across diverse real-world datasets. 
Finally, Section \ref{sec: Conclusion} concludes this article and lists some future works.

\section{Related Works and Conventional Notations}\label{sec: Related Work}
Understanding the theoretical foundations of mainstream HAD baselines is essential for developing an effective AHAD framework.
Accordingly, Section \ref{subsec: HAD} first recaps the theory and formulation of representative HAD approaches.
Then, Section \ref{subsec: HAA} briefly introduces the existing HAA methods and explains why they are not applicable to the AHAD task.
Frequently used notations are collectively defined in Section \ref{sec:notation}.

\subsection{Hyperspectral Anomaly Detection}\label{subsec: HAD}
In the RS field, HAD methods assign an anomaly score within the range of $[0,1]$ to each pixel (i.e., spectral signature) in the target HSI. 
It is worth noting that the resulting scores represent ``soft'' output rather than conventional binary classifications.
This is because the so-called anomaly is identified with respect to the surrounding background \cite{TGFA-AD,Auto-AD} and lacks an explicit, generally accepted definition. 
%

Due to the scarcity of large-scale HAD datasets with annotations, most mainstream HAD methods are developed in an unsupervised manner, commonly using a background reconstruction strategy \cite{PSSIN_Doi1,DWSDiff_Doi2,RBRSE_Doi3}.
Let $\bX\in\mathbb{R}^{C\times L}$ denote the matrix representation of an HSI with $L$ pixels, based on which we briefly introduce the mathematical formulations of HAD.
First, the robust principal component analysis (RPCA) criterion \cite{RPCA} is widely adopted to preliminarily formulate the HAD problem, 
\begin{equation}\label{eq: HAD1}
    \min_{\bB,\,\bA} \ \|\bB\|_* + \lambda \|\bA\|_1
    \quad \text{s.t.} \quad \bX = \bB + \bA,
\end{equation}
where $\bB$ denotes a low-rank background and $\bA$ represents the sparse anomaly component \cite{yao2022hyperspectral}.
Consequently, the resulting detection map can be naturally obtained as $\sum_{i=1}^C [\bA]_{i,:}$, followed by an intensity standardization.
Based on the basic criterion in \eqref{eq: HAD1}, various regularizations (e.g., confidence-weighted prior \cite{11478796} and Mahalanobis regularization \cite{xu2018joint}) have been incorporated, leading to numerous RPCA variants for the HAD task.
In particular, SuperRPCA \cite{SuperRPCA} introduces a novel collaborative superpixel representation prior into the RPCA formulation.
SuperRPCA first employs a real-time superpixel-based collaborative representation to reconstruct the rough background component.
The convex $\bQ$-quadratic norm \cite{CODE} is then incorporated to extract informative structures from the rough background component, thereby providing effective regularization for RPCA.
Nevertheless, complex and diverse real-world scenarios render the decomposition of $\bB$ and $\bA$ highly ill-posed, significantly hindering the HAD performance \cite{SlowVaryingAD_Doi5}.

Instead of further complicating the formulations (e.g., more regularizers), an effective improvement is to introduce a customized background subspace $\bD=[\bd_1,\bd_2,\cdots,\bd_k]\in\mathbb{R}^{C\times k}$.
When the subspace, typically known as a dictionary, adequately represents the background (i.e., $\bB=\bD\bS$), the residual component naturally remains only the desired anomaly matrix \cite{SSEAD_Doi4}. 
The strategy results in a more general HAD formulation, i.e.,
\begin{equation}\label{eq: LRR}
    \min_{\bS,\,\bA} \ \text{rank}(\bS) + \lambda ~\text{sparse}(\bA)
    \quad \text{s.t.} \quad \bX = \bD\bS + \bA,
\end{equation} 
which is referred to as low-rank representation (LRR) \cite{LRASR}.
Similar to the RPCA-based methods, diverse and innovative regularizations (e.g., graph and total variation \cite{GTVLRR}, deep plug-and-play \cite{DeCNN}), customized dictionary construction \cite{ADLR,PABDC}, and the tensor-based LRR \cite{PTA,LARTVAD} have been further developed to improve the HAD performance. 
On the other hand, integrating subspace-guided strategies with a transformer is also proposed for HAD (e.g., TGFA-AD \cite{TGFA-AD}), which first projects the target HSI into the discriminative abundance domain via convex geometry \cite{HyperCSI}, followed by applying a transformer-guided fractional attention to capture the anomalies.

Recently, both quantum deep networks (QUEEN) \cite{HyperQUEEN,HyperKING} and classical neural networks \cite{9449622,SpectralGPT} have achieved remarkable success in the RS field.
As a representative example, HyFuHAD \cite{HyFuHAD} integrates Einstein fuzzy computing with QUEEN to develop a quantum-classical fuzzy multi-criteria decision-making (MCDM) framework for HAD.
By exploiting unique and informative quantum features, the QUEEN-based MCDM can substantially enhance the overall detection performance \cite[Section IV-C]{HyFuHAD}.
We remark that due to the scarcity of large-scale annotated data, unsupervised and single-data learning have become the mainstream strategies for developing DL-based HAD methods.
For instance, following a strategy similar to the well-known deep image prior (DIP) \cite{DIP,GQmu}, a fully convolutional neural network (CNN) \cite{Auto-AD} can be used to reconstruct background structures $\bB$ from a randomly sampled noise tensor.
Under an unsupervised reconstruction, the distance between the reconstruction and the original target HSI is typically employed as an optimization objective \cite{TGFA-AD}.
Since the background component dominates the overall HSI structure (i.e., $\bB \approx\bX$), such an unsupervised scheme can effectively recover $\bB$, with the residual component treated as the anomaly (i.e.,  $\bA=\bX-\bB$).
Building upon this unsupervised strategy, customized architectural design (e.g., blind-spot network \cite{BockNet, PUNNet}, gated transformer \cite{TGFA-AD}, fractional attention \cite{TGFA-AD}) and regularization strategies (e.g., graph structure \cite{RGAE} and adaptive weighted loss \cite{Auto-AD}) have been proposed to further suppress the anomaly effect during the reconstruction.

According to the above perspectives, directly adopting a minus anomaly component $-\bA$ as the perturbation signal $\mathbf{P}$ appears to be a reasonable and straightforward approach for anomaly assimilation. 
However, in practice, HAD baselines only need to ensure that anomalous pixels exhibit a relatively large disparity in $\bA$, rather than completely decompose anomaly components. 
As a result, the energy of $\bA$ is extremely small and leads to a negligible impact on the overall hyperspectral structure.
In other words, even under the perfect CSI assumption (i.e., zero-pixel shift), the detection maps obtained using $\bX$ and $\bX-\bA$ remain almost the same [{see} {Fig.}~\ref{fig: intro_fig}(c)], indicating the need for a fundamentally new AHAD solution.

\subsection{Hyperspectral Adversarial Attack}\label{subsec: HAA}
Recently, DL models for vision tasks, such as classification and object recognition, have demonstrated their remarkable capabilities.
However, these advanced techniques remain vulnerable to small deviations.
In particular, the hierarchical structures of deep networks may amplify imperceptible deviations \cite{cisse2017parseval}, leading to substantially biased predictions \cite{akhtar2021advances}. 
For instance, a deep classifier may misclassify a dog image as a bird category when subjected to certain designed perturbations.  
Such signals are referred to as adversarial perturbations, and the techniques for generating them are known as adversarial attacks (AAs). 
Among existing AA methods, gradient-based approaches are widely studied and are generally regarded as particularly effective \cite{yuan2021meta}. 
Given an input image $\bx$ and its corresponding reference label $\by$, gradient-based attacks estimate the perturbation signal $\boldsymbol{\delta}$ by performing a single-step or iterative optimization via  
\begin{equation}\label{eq: AA}
\boldsymbol{\delta}^{\star}=\arg\max_{\|\boldsymbol{\delta}\|_p \leq \epsilon} \; \mathcal{L}(f_\theta(\bx + \boldsymbol{\delta}), \by),
\end{equation}
where $f_\theta(\cdot)$ denotes the target network with pretrained parameters $\theta$, while $\mathcal{L}(\cdot)$ and $\epsilon$ represent the task-specific objective (e.g., cross-entropy for classification) and a predefined attack budget in terms of $\ell_p$-norm. 
Representative gradient-based AA methods include the fast gradient sign method (FGSM) \cite{FGSM}, basic iterative method (BIM) \cite{BIM}, and projected gradient descent method (PGD) \cite{PGD}.
Notably, AA methods typically require full access to the target network (e.g., its pretrained parameters and architecture) and the reference label to generate adversarial perturbations, which are referred to as white-box attacks.
However, the information regarding the target network may be only partially available (or even unavailable) in practice.
A realistic approach is to construct a surrogate model for the target one and use it to perform AA estimations \cite{liu2016delving}. 
In light of the high transferability of adversarial perturbations \cite{papernot2016transferability}, this strategy enables AA methods to effectively attack target models using only incomplete information, a strategy known as a black-box attack.
In the hyperspectral field, AA algorithms have also been extended to material identification tasks \cite{9585480}, emerging as a security-critical research topic and attracting considerable attention. 
For example, boundary adversarial samples (BASs) are generated via a modified DeepFool scheme \cite{9585480}, where the AA loss function is iteratively optimized through network backpropagation.
However, unlike conventional 3-band RGB images, remotely sensed hyperspectral data is inherently high-dimensional but contains substantial spectral redundancy.
This poses a significant challenge for HAA because different deep classifiers may exploit distinct subsets of spectral bands for decision-making, leading to struggles with transferability and robustness.
To address this limitation, a universal object-level attack strategy \cite{10328770}, built upon a spatial-spectral superpixel template, has been developed to improve baseline HAA methods, including FGSM, momentum iterative FGSM (MI-FGSM) \cite{dong2018boosting}, and variance tuning MI-FGSM (VMI-FGSM) \cite{wang2021enhancing}.
More recently, a physically interpretable HAA method, named sparse unmixing-guided AA (SUGAA) \cite{11141475}, has been proposed using blind source separation (BSS) techniques \cite{HyperCSI}.
In the SUGAA framework, the adversarial perturbations are obtained using gradient-based methods in the low-dimensional abundance domain \cite{TGFA-AD} to ensure the physical consistency of natural HSIs, thereby improving the effectiveness and robustness of HAA.

Nevertheless, as shown in {Fig.}~\ref{fig: intro_fig}(b), such powerful techniques remain theoretically inapplicable to our AHAD task, whether in white-box or black-box settings.
This limitation stems from the fact that both AA for computer vision tasks and HAA techniques generally follow the principle in \eqref{eq: AA}, and mostly adopt gradient-based signal estimation.
Specifically, AAs are formulated to maximize the loss function of a pretrained DL model [{see} \eqref{eq: AA}], thereby effectively amplifying perturbations and eventually leading to biased results.
However, mainstream DL-based HAD detectors are generally unsupervised and require optimization from scratch for each detection ({see} Section \ref{subsec: HAD}), making AHAD fundamentally different from AA/HAA tasks. 

Under such unsupervised settings, the parameters are dynamically updated, making AA optimizations difficult and hindering their applicability to AHAD.
Moreover, even if the HAD loss function $\mathcal{L}$ (typically some reconstruction errors) can be maximized using certain AA frameworks, this does not necessarily degrade the detection performance.
In fact, the HAD performance may even be improved as long as the reconstruction errors of anomaly pixels remain relatively substantial. 
Furthermore, even if the above limitations were alleviated, the existence of state-of-the-art (SOTA) non-DL-based HAD methods would still render the AA-based AHAD ineffective.
The above limitations, collectively, highlight the necessity of developing a radically new theory for AHAD.

\subsection{Notations}\label{sec:notation}

The notations used throughout this article are defined below.
Throughout this article, we use regular letters $z$, lowercase bold italic letters $\bz$, uppercase bold italic letters $\bZ$, and uppercase bold upright letters $\mathbf{Z}$ to denote scalar, vector, matrix, and tensor, respectively.
Given a positive integer $K$, $\mathcal{I}_K\triangleq\{1\cdots,K\}$.
$\mathbb{R}^{M\times N}$ and $\mathbb{R}^{M\times N\times C}$ denote real-valued $(M\times N)$-dimensional matrix space and $(M\times N\times C)$-dimensional tensor space, respectively.
%
%
Given a $M$-way tensor $\mathbf{X}$, $[\mathbf{X}]_{a_1,a_2,\cdots,a_M}$ and $\mathbf{X}^{(n)}$ denotes its $(a_1,a_2,\cdots,a_M)$th entry and its mode-$n$ matricization \cite{GQmu}, respectively.
$max(\mathbf{X})$ refers to the absolute maximum entry of the tensor $\mathbf{X}$.
Moreover, given a matrix $\bZ$ with appropriate dimension, the tensor mode-$n$ multiplication \cite{GQmu} between $\mathbf{X}$ and $\bZ$ is denoted as $\mathbf{X}\times_n\bZ$.
$\|\cdot\|_*$, $\|\cdot\|_1$, $\|\cdot\|_F$, and $\odot$ denote the nuclear norm, $\ell_1$-norm, Frobenius norm, and elementwise multiplication, respectively.
For the tensor $\ell_1$-norm and Frobenius norm, we use the same notation as for the matrix, while the explicit definitions are provided upon first use.
$\nabla_{x}$ represents the discrete differential operator \cite{GTVLRR} along the $x$ direction.

\section{Proposed Unsupervised AHAD Method}\label{sec: Proposed}
To the best of our knowledge, this study presents the first AHAD framework in the RS field.
To explicitly define the AHAD task, we first present a detailed problem definition in Section \ref{subsec: perfect CSI Formulation}.
Accordingly, an AHAD criterion with perfect CSI is proposed and solved in Section \ref{sec:perfectCSI}, where a novel ARAB regularization is customized to address the challenging AHAD task.
To facilitate practical applicability, the uncertainty induced by imperfect CSI is further described as matrix-shifting misalignment, motivating us to design a robust probabilistic AHAD criterion, as detailed in Section \ref{subsec: Robust}.
Implementation details are presented in Section \ref{subsec: Implementation}.

\subsection{Problem Characteristics of Anti-HAD}\label{subsec: perfect CSI Formulation}

We define the target AHAD problem along with several practical considerations. 
For a target area requiring anti-detection, let $\mathbf{Y}_H\in\mathbb{R}^{H\times W\times C}$ denote the corresponding HSI remotely sensed by an unknown hyperspectral detector ({see} {Fig.}~\ref{fig: AHAD_concept}).
Here, $H$, $W$, and $C$ represent the image height, image width, and number of spectral bands, respectively.    
The AHAD problem can be characterized as follows:
\begin{enumerate}
\item[(C1)]
\textit{(Black-Box Uncertainty)}
Let $\mathcal{A}(\cdot)$ denote an arbitrary HAD method from the opponent's side ({see} {Fig.}~\ref{fig: AHAD_concept}), which is assumed to be unknown yet effective.

\smallskip

\item[(C2)]
\textit{(ARAB Property)}
The goal of AHAD is to estimate a perturbation/anti-detection signal $\mathbf{P}\in\mathbb{R}^{H\times W\times C}$, such that the perturbed HSI (i.e., $\mathbf{Y}_A=\mathbf{Y}_H+\mathbf{P}$) yields an anomaly detection map $\mathcal{A}(\mathbf{Y}_A)$ in which the ground facilities and activities (of our side) are barely identified ({see} {Fig.}~\ref{fig: AHAD_concept}).

\smallskip

\item[(C3)]
\textit{(Energy Efficiency, EE)}
Minimum energy consumption of $\mathbf{P}$ is essential to control the transmit power \cite{8335785,qiao2007interference}.   
Since the transmission of the anti-detection signal relies on the base station ({see} {Fig.}~\ref{fig: AHAD_concept}), unbounded or excessive transmitting power remains impractical in real-world scenarios.
Thus, as commonly considered in wireless communications \cite{chi2017convex}, minimizing the energy of the anti-detection signal $\mathbf{P}$ while fulfilling the AHAD requirement (C2) leads to a more practical solution.

\smallskip

\item[(C4)]
\textit{(Restoration Resistance, RR)}
Surely, for reliable anti-detection, the perturbation $\mathbf{P}$ should be robust against the removal/restoration techniques $\mathcal{D}(\cdot)$.
Specifically, noise removal and restoration techniques, which can recover clean HSIs from severely corrupted observations, have been widely used as preprocessing steps in the recent RS field \cite{zhang2026real,11345127,9318503}.
For instance, advanced quantum techniques can effectively remove mixed noise (i.e., Gaussian, impulse, and stripe noise) for NASA's highly damaged HSI \cite{HyperQUEEN}, enabling subsequent sophisticated blind signal processing tasks \cite{HyperKING}.
Such advanced techniques suppress various noise types, so noise-like anti-detection signals $\mathbf{P}$ can not reliably achieve AHAD [{see} {Fig.}~\ref{fig: intro_fig}(a)].
Therefore, an effective anti-detection signal $\mathbf{P}$ should not be recognized as noise effect by the denoiser $\mathcal{D}(\cdot)$ (i.e., $\mathcal{D}(\mathbf{Y}_A)\approx\mathbf{Y}_A$).
\end{enumerate}
Note that simultaneously achieving (C3) and (C4) already seems infeasible, as they appear to contradict each other, not to mention the other two challenging characteristics, (C1)-(C2), and additional real-world considerations (e.g., imperfect CSI, to be discussed next).
Existing HAA techniques hence remain inapplicable for AHAD, whether under white-box or black-box settings ({see} Section \ref{subsec: HAA}).

Under (C1)-(C4), we will develop an AHAD criterion with perfect CSI information ({see} Section \ref{sec:perfectCSI}), and then extend it to a robust probabilistic criterion to address the uncertainty of CSI ({see} Section \ref{subsec: Robust}).
To this end, we first discuss the definitions of perfect/imperfect CSI.

The remotely sensed HSI $\mathbf{Y}_H$ is essential for estimating $\mathbf{P}$, but it is typically unavailable.
In practice, however, the challenge in the availability of $\mathbf{Y}_H$ can be reduced to addressing its spatial misalignment caused by imperfect CSI (e.g., imperfectly estimated detector position).
In this case, a \textit{surrogate} HSI $\widetilde{\mathbf{Y}}_H$ can be reasonably obtained to approximate $\mathbf{Y}_H$ when estimating the anti-detection signal $\mathbf{P}$.
\begin{Assumption}\label{Assumption 1}
Through some high-confidence CSI estimation, the surrogate HSI $\widetilde{\mathbf{Y}}_H$ can reasonably be obtained to serve as an alternative to the unknown $\mathbf{Y}_H$.
\hfill$\square$
\end{Assumption}
Specifically, the position of the unknown detector can be effectively estimated based on contemporary techniques, such as radar-based localization \cite{uav_detection_1}.
With this essential yet obtainable information, powerful AI-empowered imaging systems (e.g., sensor pose estimation \cite{shetty2019uav} and viewing image synthesis \cite{li2026unifying,vuong2025aerialmegadepth}) can exploit different geometric perspectives (e.g., our own sensor) along with estimated detector position to synthesize the corresponding images $\widetilde{\mathbf{Y}}_H$.
Even when synthesized images exhibit limited spectral resolutions (e.g., multispectral images), advanced spectral super-resolution algorithms enable the reconstruction of a high-fidelity HSI from its multispectral counterpart \cite{COS2A,young2026spectral}.
Therefore, the challenge in the availability of $\mathbf{Y}_H$ can be greatly alleviated, as its surrogate $\widetilde{\mathbf{Y}}_H$ can be reasonably obtained based on contemporary technologies.
Even though, the estimated detector CSI could still remain imperfect; thus, directly assuming $\widetilde{\mathbf{Y}}_H=\mathbf{Y}_H$ may be overly idealized.
In fact, images acquired from different positions naturally exhibit spatial misalignment.
Therefore, when estimating the anti-detection signal based on $\widetilde{\mathbf{Y}}_H$, an AHAD criterion robust to the spatial misalignment between $\widetilde{\mathbf{Y}}_H$ and true $\mathbf{Y}_H$ (i.e., pixel shifting) should be considered to facilitate reliable anti-detection, as detailed in Section \ref{subsec: Robust}.

\subsection{AHAD Criterion for Perfect CSI}\label{sec:perfectCSI}

According to (C1)-(C4), we first design the AHAD criterion for the zero-shifting case.
Let $\widetilde{\mathbf{Y}}_A=\widetilde{\mathbf{Y}}_ H+\mathbf{P}$ denote the perturbed HSI with respect to the surrogate HSI $\widetilde{\mathbf{Y}}_H$.
Generally, the AHAD criterion can be cast as the following optimization problem, i.e.,
\begin{align}\label{eq: attack_overall}
\min_{\mathbf{P}} &\ \|\mathbf{P}\|_F +
\text{REG}_1(\mathbf{P})+\text{REG}_2(\widetilde{\mathbf{Y}}_A) \quad  \notag
\\
&
 \text{s.t.} \quad\widetilde{\mathbf{Y}}_A \in [0,1]^{H\times W \times C},
\end{align}
where $\|\mathbf{P}\|_F\triangleq\sqrt{\sum_{i=1}^H\sum_{j=1}^W\sum_{k=1}^C [\mathbf{P}]_{i,j,k}^2}$ represents the energy of the anti-detection signal in terms of the tensor Frobenius norm. 
%
This energy term $\|\mathbf{P}\|_F$, together with the box constraint (i.e., $\widetilde{\mathbf{Y}}_A \in [0,1]^{H\times W \times C}$), reduces energy consumption and yields a practically feasible solution, thereby satisfying the EE requirement in (C3).
In addition, $\text{REG}_1(\cdot)$ and $\text{REG}_2(\cdot)$ are used to regularize the anti-detection signal and perturbed HSI, respectively.
In the AHAD criterion \eqref{eq: attack_overall}, $\text{REG}_1(\cdot)$ enforces the RR in (C4), while $\text{REG}_2(\cdot)$ encourages the ARAB property in (C2).
Comprehensive experiments ({see} Section \ref{sec: Experiment}) will demonstrate that the single anti-detection signal $\mathbf{P}$ (computed by \eqref{eq: attack_overall}) can substantially degrade the performance of most benchmark HAD methods, without requiring the prior information of detectors, thereby addressing the black-box uncertainty in (C1). 
Next, we explicitly design the two regularizers. 

For the first regularizer $\text{REG}_1$, we employ the spectral-spatial total variation (SSTV), defined as
\begin{align}\label{eq: SSTV}
    \text{REG}_1(\mathbf{P})
    &:= \text{SSTV}(\mathbf{P}) \nonumber\\
    &\triangleq 
    \alpha \|\nabla_{H}\mathbf{P}\|_1 
    + \beta \|\nabla_{W}\mathbf{P}\|_1 
    + \gamma \|\nabla_{C}\mathbf{P}\|_1,
\end{align}
where $\|\mathbf{X}\|_1\triangleq{\sum_{i=1}^H\sum_{j=1}^W\sum_{k=1}^C |[\mathbf{X}]_{i,j,k}|}$ is the tensor $\ell_1$-norm of $\mathbf{X}\in\mathbb{R}^{H\times W\times C}$.
As discussed in (C4), a reliable anti-detection signal should not be recognized as a noise effect.
In general, noise or outlier effects tend to exhibit noticeable and unfavorable non-smoothness \cite{zhang2026real,11345127}.
The SSTV promotes the smoothness of $\mathbf{P}$, thereby preventing the AHAD criterion from producing a high-frequency, non-smooth solution. 
See our discussions in Section \ref{subsec: Ablation}.

Next, we design the ARAB regularizer, $\text{REG}_2$, to encourage $\widetilde{\mathbf{Y}}_A$ to meet (C2).
Recall that hyperspectral anomalies are pixels whose spectral characteristics deviate significantly from those of the surrounding background \cite{TGFA-AD}.
Moreover, the background components in HSIs are typically composed of large and smooth regions.
According to these observations, the anomaly effect is likely to induce spatial discontinuities, thereby increasing the variation in HSIs relative to their anomaly-assimilated counterparts (i.e., the same HSIs but removing anomalies).  
To achieve (C2), the ARAB regularization can be mathematically formulated by flattening the spatial variation of the perturbed HSI, which is to minimize
\begin{align}\label{eq: variation 1}
    \|\nabla_{H}\widetilde{\mathbf{Y}}_A\|_F+\|\nabla_{W}\widetilde{\mathbf{Y}}_A\|_F.
\end{align}
Here, we adopt the Frobenius norm rather than the commonly used $\ell_1$-norm in TV, since the latter promotes gradient sparsity and concentrates variation in a specific region, which may instead strengthen the anomaly effect rather than promote the ARAB property.
By contrast, \eqref{eq: variation 1} is actually the ridge regression that promotes uniformly distributed gradients, thereby having the effect to flatten the spatial variation of $\widetilde{\mathbf{Y}}_A$.

However, minimizing variation across high-dimensional (i.e., $C$-band) hyperspectral signatures may reduce the efficiency of the ARAB process due to spectral redundancy.
Fortunately, HSIs typically lie in an $N$-dimensional subspace ($N\ll C$) \cite{HyperQUEEN,SISHY}.
This motivates us to flatten the topology of the perturbed data in the feature space, obtained by the affine set fitting (ASF) \cite{HyperCSI,ASF}.
Specifically, we first center $\widetilde{\mathbf{Y}}_A$ at the origin by mean shifting, i.e., $\widetilde{\mathbf{Y}}_A^s=\widetilde{\mathbf{Y}}_A-\widetilde{\mathbf{M}}_A$, where $\widetilde{\mathbf{M}}_A$ denotes the mean tensor of $\widetilde{\mathbf{Y}}_A$, defined as
\[
[\widetilde{\mathbf{M}}_A]_{i,j,:}=\frac{1}{HW}\sum_{i=1}^H\sum_{j=1}^W [\widetilde{\mathbf{Y}}_A]_{i,j,:},\quad\forall(i,j)~\in\mathcal{I}_H\times\mathcal{I}_W.
\]
The mean-shifted HSI $\widetilde{\mathbf{Y}}_A^s$ is then projected onto an $N$-dimensional affine set to obtain the feature tensor, i.e.,
\[
\widetilde{\mathbf{Z}}_A=\widetilde{\mathbf{Y}}_A^s\times_3\bU 
\in\mathbb{R}^{H\times W\times N},
\]
where $\bU\in\mathbb{R}^{C\times N}$ consists of the top-$N$ left singular vectors of the mode-$3$ unfolded mean-shifted HSI $\widetilde{\mathbf{Y}}_A^{s(3)}$.
This allows us to reformulate \eqref{eq: variation 1} to a more efficient form, i.e.,
\begin{align}\label{eq: variation 2}
    \|\nabla_{H}\widetilde{\mathbf{Z}}_A\|_F+\|\nabla_{W}\widetilde{\mathbf{Z}}_A\|_F,
\end{align}
which is a function of $\mathbf{P}$.
In addition to dimensionality reduction, a topology enhancement strategy is developed to strengthen the ARAB process on the real anomalies.
In particular, before performing anti-detection, we observe that anomalies within $\mathbf{Y}_H$ (resp., $\widetilde{\mathbf{Y}}_H$) often distribute relatively far from the data mean in the feature space.
This motivates us to construct a D2CM $\mathbf{W}\in\mathbb{R}^{H\times W \times N}$ by
\begin{align}\label{eq: define W}
\mathbf{W}:=\eta\left(\widetilde{\mathbf{Z}}_H\odot\widetilde{\mathbf{Z}}_H\right),
\end{align}
where $\widetilde{\mathbf{Z}}_H$ can be obtained by performing the ASF on the surrogate $\widetilde{\mathbf{Y}}_H$.
Additionally, the scaling factor $\eta\geq0$ determines the strength of this topology enhancement strategy.
More specifically, based on the manifold assumption \cite{GTVLRR}, in the low-dimensional subspace, data points deviating from the background manifold are commonly referred to as anomalies \cite{GTVLRR}.
Nevertheless, the absence of annotations makes it difficult to identify the background manifold, hence hindering the use of the standard point-to-manifold distance (P2M) \cite{burago2001course} for the topology enhancement.
Alternatively, we employ the D2C as the surrogate distance of the standard P2M.
Specifically, background components dominate the overall data structure, whereas anomalies account for only a small proportion of HSI.
The overall data mean is expected to be very close to the background mean, which corresponds to the role of the center of the background manifold.
From this perspective, the entries of D2CM $\mathbf{W}$ can be regarded as a preliminary yet straightforward measure of the deviation from the anomaly to the background manifold.

Accordingly, $\mathbf{W}$ can be used to radially scale each data point of $\widetilde{\mathbf{Z}}_A$, i.e., $\widetilde{\mathbf{Z}}_A\odot\mathbf{W}$.
This topology-enhancement strategy further amplifies the variation introduced by real anomalies.
If the ARAB process makes anomalies indistinct in the topology-enhanced $\widetilde{\mathbf{Z}}_A\odot\mathbf{W}$, the anomalies in $\widetilde{\mathbf{Z}}_A$ should also be well assimilated into the background, indicating a highly effective ARAB procedure.
Consequently, the topology-enhanced ARAB regularization with respect to the weight $\mathbf{W}$ can be explicitly expressed as
\begin{align}\label{eq: topology-enhanced ARAB}
    \Phi_{\mathbf{W}}(\widetilde{\mathbf{Z}}_A)\triangleq& \sum_{i=1}^N \|\nabla_{H}[\widetilde{\mathbf{Z}}_A\odot \mathbf{W}]_{:,:,i}\|_F
    +
    \|\nabla_{W}[\widetilde{\mathbf{Z}}_A\odot \mathbf{W}]_{:,:,i}\|_F.
\end{align}
Once the topology-enhanced feature $\widetilde{\mathbf{Z}}_A\odot \mathbf{W}$ exhibits sufficiently limited variations, the original $\widetilde{\mathbf{Y}}_A$ naturally fulfills the AHAD purpose.
By \eqref{eq: topology-enhanced ARAB}, we term the resulting anti-detection signals Lipschitz-forcing perturbations, as the topology-enhanced ARAB regularization strongly suppresses local pixel-wise variations (i.e., discrete gradient).
This echoes the Lipschitz constraint bounding the DL network sensitivity (i.e., network gradient) and facilitates stable models \cite{9319198}.

Finally, to further fool the unknown detectors [{see} (C1)], we introduce the tail energy-based regularization to produce a pseudo-anomaly effect.
Although the anomaly effect can be significantly suppressed by \eqref{eq: topology-enhanced ARAB}, advanced detectors may still be able to detect true anomalies owing to their high sensitivity to the anomalous signatures.
Therefore, a straightforward approach is to generate pseudo-anomalies on the background regions, serving as part of the ARAB process, as the resulting pseudo-backgrounds (i.e., pseudo-anomalies corrupted backgrounds) also help assimilate the real anomalies, thereby further strengthening the ARAB regularization.

This goal can be mathematically formulated as to maximizing the tail singular values (SVs) of the perturbed HSI. 
Specifically, after removing the global mean spectrum, the background components usually dominate the overall data structure of the mean-shifted HSI, which typically lies in some low-dimensional subspace.
These observations suggest that most of the background energy concentrates along the top-$N$ principal directions (PDirs), which can be characterized by the top-$N$ SVs, i.e., $\sigma_1 \ge \sigma_2 \ge \cdots \ge \sigma_N$. %
In contrast, since anomalies deviate from the background components, they are unlikely to share consistent PDirs with the background.
Hence, their energy is expected to be retained along the remaining tail PDirs.
Consequently, by Eckart–Young–Mirsky theorem \cite{golub1987generalization}, the pseudo-anomaly generation of $\widetilde{\mathbf{Y}}_A$ can be mathematically formulated by maximizing the tail energy, i.e.,
\begin{align}\label{eq: pseudo-anomalies}
    \|\widetilde{\mathbf{Y}}_A^s-\widetilde{\mathbf{Y}}_A^s\times_3\bU^T\times_3\bU\|_F.
\end{align}
Remarkably, although maximizing the tail SVs may also induce noise effects, the usage of SSTV regularization [{see} \eqref{eq: SSTV}] can effectively mitigate this dilemma when introducing the pseudo-anomaly effect. 
Thus, we conclude that the designs of $\text{REG}_1(\cdot)$ and $\text{REG}_2(\cdot)$ echo each other.
Overall, given $\lambda_1, \lambda_2\geq0$ as the trade-off parameters, the ARAB regularization $\text{REG}_2(\widetilde{\mathbf{Y}}_A)$ can be expressed as
\begin{align}\label{eq: REG2}
    \text{REG}_2(\widetilde{\mathbf{Y}}_A):=~& \lambda_1 \Phi_{\mathbf{W}}(\widetilde{\mathbf{Z}}_A)-\lambda_2\|\widetilde{\mathbf{Y}}_A^s-\widetilde{\mathbf{Y}}_A^s\times_3\bU^T\times_3\bU\|_F,
\end{align}
wherein $\widetilde{\mathbf{Y}}_A^s$, $\bU$, and $\widetilde{\mathbf{Z}}_A$ are derivable from $\widetilde{\mathbf{Y}}_A$.
Therefore, we have completed the design of the AHAD criterion \eqref{eq: attack_overall}, whose implementation will be discussed in Section \ref{subsec: Implementation}.

\subsection{Robust AHAD Criterion for Imperfect CSI}\label{subsec: Robust}
Since the CSIs (e.g., detector position) require additional estimation ({see} Assumption \ref{Assumption 1}), they remain inherently imperfect.
In practice, images acquired from different viewpoints naturally exhibit spatial misalignment.
Accordingly, we further extend \eqref{eq: attack_overall} to a robust AHAD criterion to address unknown misalignment (pixel-shifting) between $\widetilde{\mathbf{Y}}_A$ and $\mathbf{Y}_A$ for reliable anti-detection.
To this end, we first elaborate on the modeling of the pixel-shiftings, in order to mathematically define the spatial misalignment between $\widetilde{\mathbf{Y}}_A$ and $\mathbf{Y}_A$.

In real-world scenarios, the internal geometrical calibration of high-quality airborne hyperspectral imagery can achieve sub-pixel-level accuracy \cite{green1998imaging}.
However, the spatial misalignment caused by imperfect CSI is more similar to that in cross-sensor cases (e.g., RS image fusion \cite{10543175}), which are commonly modeled as pixel-level shifting in the RS field \cite{11435906}.
Hence, let $r$ be a non-negative integer denoting the considered maximum pixel-shifting range.
The collection of all candidate $(2r+1)^2$ pixel-shifting cases can be written as
\begin{equation}\label{eq: S}
\mathcal{S}=\{-r,\ldots,+r\}\times\{-r,\ldots,+r\}.
\end{equation}
Specifically, for a pixel-shifting case $\boldsymbol{\Delta}=(\Delta_h,\Delta_w)\in\mathcal{S}$, the pixel-shifted HSI $\mathcal{T}_{\boldsymbol{\Delta}}(\widetilde{\mathbf{Y}}_H)\in\mathbb{R}^{H\times W\times C}$ can be explicitly written as
\begin{equation}\label{eq: shifting}
[\mathcal{T}_{\boldsymbol{\Delta}}(\widetilde{\mathbf{Y}}_H)]_{h,w,:}
=
[\widetilde{\mathbf{Y}}_H]_{(h+\Delta_h),(w+\Delta_w),:},
\end{equation}
where the boundary pixels can be determined based on a predefined boundary rule, such as padding, cropping, or resampling ({see} Section \ref{subsec: exp setting}). 
As boundary pixels account for a relatively small proportion of the HSI, they are likely to have a limited effect on the subsequent optimization.
Accordingly, $\mathbf{Y}_H$ and $\widetilde{\mathbf{Y}}_H$ are said to have a pixel-shifting $\boldsymbol{\Delta}$ if $\mathbf{Y}_H=\mathcal{T}_{\boldsymbol{\Delta}}(\widetilde{\mathbf{Y}}_H)$ under an appropriate boundary rule.
This definition also covers the perfect CSI case [i.e., $\boldsymbol{\Delta}=(0,0)$] as it naturally yields $\mathbf{Y}_H=\mathcal{T}_{(0,0)}(\widetilde{\mathbf{Y}}_H)=\widetilde{\mathbf{Y}}_H$.

Based on the above definitions, we proceed to the design of the robust AHAD criterion.
Let $\mathcal{L}(\mathbf{P}|\widetilde{\mathbf{Y}}_H)= \|\mathbf{P}\|_F +\text{REG}_1(\mathbf{P})+\text{REG}_2(\widetilde{\mathbf{Y}}_A)$ denote the objective function of AHAD [{see} \eqref{eq: attack_overall}] conditioned on the surrogate HSI.
If the pixel-shifting $\boldsymbol{\Delta}$ is known ({see} {Fig.}~\ref{fig: AHAD_concept}), the AHAD criterion in \eqref{eq: attack_overall} can be readily reformulated as
\begin{align}\label{eq: single shift}
     \min_{\mathbf{P}}\quad 
\mathcal{L}(\mathbf{P}\mid \mathcal{T}_{\boldsymbol{\Delta}}(\widetilde{\mathbf{Y}}_H))\quad \mathrm{s.t.}\quad 
\widetilde{\mathbf{Y}}_A^{(\boldsymbol{\Delta})}
\in [0,1]^{H\times W\times C},
\end{align}
where $\widetilde{\mathbf{Y}}_A^{(\boldsymbol{\Delta})}\triangleq \mathcal{T}_{\boldsymbol{\Delta}}(\widetilde{\mathbf{Y}}_H)+\mathbf{P}$ denotes the perturbed HSI with respect to the pixel-shifted HSI, i.e., $\mathcal{T}_{\boldsymbol{\Delta}}(\widetilde{\mathbf{Y}}_H)$.
In practice, however, the exact pixel-shifting is typically unknown because of CSI uncertainty. 
In other words, directly optimizing \eqref{eq: single shift}, conditioned on a single-shifting case, is unlikely to yield a robust anti-detection signal.
Furthermore, although such uncertainty may be addressed by considering an expectation-based AHAD criterion, such as $\mathbb{E}_{\boldsymbol{\Delta}\sim p}[\mathcal{L}(\mathbf{P}|\mathcal{T}_{\boldsymbol{\Delta}}(\widetilde{\mathbf{Y}}_H))]$ ({see} Appendix \ref{sec:proof Lemma 1}), it is not straightforward to optimize the probabilistic criterion.
The lemma addresses this challenge:
\begin{Lemma}\label{Lemma 1}
Under the imperfect CSI scenario, the induced robust probabilistic AHAD criterion can be cast as the deterministic optimization, i.e.,   
\begin{align}
\label{eq: robust AHAD}
  \min_{\mathbf{P}}\quad 
& \sum_{\boldsymbol{\Delta}\in\mathcal{S}}
q^{(\boldsymbol{\Delta})}
\mathcal{L}\!\left(
\mathbf{P}
\mid
\mathcal{T}_{\boldsymbol{\Delta}}(\widetilde{\mathbf{Y}}_H)
\right)\notag
\\
\mathrm{s.t.}\quad 
& \widetilde{\mathbf{Y}}_A^{(\boldsymbol{\Delta})}
\in [0,1]^{H\times W\times C},\quad\forall\boldsymbol{\Delta}\in\mathcal{S},
\end{align}
for some convex profile, $q^{(\boldsymbol{\Delta})}\geq 0$, $\sum_{\boldsymbol{\Delta}\in\mathcal{S}}q^{(\boldsymbol{\Delta})}=1$, denoting the significance of each pixel-shifting $\boldsymbol{\Delta}$.\hfill$\square$
\end{Lemma}
The deterministic form in Lemma \ref{Lemma 1} facilitates the subsequent implementation.
The probabilistic AHAD criterion and the proof of Lemma \ref{Lemma 1} are relegated to Appendix \ref{sec:proof Lemma 1}.

To complete the robust anti-detection signal $\mathbf{P}$, we practically specify $q^{(\boldsymbol{\Delta})}$ before implementing \eqref{eq: robust AHAD}.
To this end, several discussions regarding the pixel-shifting are presented below.
First, $\widetilde{\mathbf{Y}}_H$ is assumed to be obtainable through some high-confidence CSI estimation (i.e., Assumption \ref{Assumption 1}).
Building upon this, $q^{(\boldsymbol{\Delta})}|_{\boldsymbol{\Delta}:=(0,0)}$ is naturally assigned with the highest weight/significance, as it directly corresponds to the AHAD objective conditioned on $\widetilde{\mathbf{Y}}_H$ [{see} \eqref{eq: attack_overall}], leading to requirement (R1).
Next, when no prior information on the shifting direction is available, assuming a direction-agnostic $q^{(\boldsymbol{\Delta})}$ is expected to improve the generalizability.
In other words, the assigned weight depends only on the magnitude of $\boldsymbol{\Delta}$, leading to requirement (R2).
Finally, under high-confidence CSI estimation, a more severe pixel-shifting should naturally correspond to a lower contribution.
Together with (R2), $q^{(\boldsymbol{\Delta})}$ is expected to be inversely proportional to the magnitude $||\boldsymbol{\Delta}||_2$, leading to requirement (R3).
We mathematically summarize the three premises for defining a reasonable $\{q^{(\boldsymbol{\Delta})}\}$, i.e.,
\begin{equation}\label{eq: require}
\left\{
\begin{aligned}
\text{(R1)}\quad 
& q^{(\mathbf{0})} \geq q^{(\boldsymbol{\Delta})},
\quad \forall \boldsymbol{\Delta}\in\mathcal{S}; \\[1mm]
\text{(R2)}\quad 
& \|\boldsymbol{\Delta}_i\|_2 = \|\boldsymbol{\Delta}_j\|_2
\Rightarrow 
q^{(\boldsymbol{\Delta}_i)}=q^{(\boldsymbol{\Delta}_j)}, \\
& \hspace{1em} \forall \boldsymbol{\Delta}_i,\boldsymbol{\Delta}_j\in\mathcal{S}; \\[1mm]
\text{(R3)}\quad 
& \|\boldsymbol{\Delta}_i\|_2 < \|\boldsymbol{\Delta}_j\|_2 
\Rightarrow 
q^{(\boldsymbol{\Delta}_i)}>q^{(\boldsymbol{\Delta}_j)}, \\
& \hspace{1em} \forall \boldsymbol{\Delta}_i,\boldsymbol{\Delta}_j\in\mathcal{S}.
\end{aligned}
\right.
\end{equation}
Obviously, the weights in Gaussian kernel satisfy (R1)-(R3), as well as the non-negative and sum-to-one constraints in Lemma \ref{Lemma 1}, thereby leading to the natural definition of $q^{(\boldsymbol{\Delta})}$, i.e.,
\begin{equation}
q^{(\boldsymbol{\Delta})}
=
\frac{
\exp\!\left(
-\frac{\|\boldsymbol{\Delta}\|_2^2}{2\sigma^2}
\right)
}{
\sum_{\boldsymbol{\Delta}'\in\mathcal{S}}
\exp\!\left(
-\frac{\|\boldsymbol{\Delta}'\|_2^2}{2\sigma^2}
\right)
},
\quad
\forall\boldsymbol{\Delta}\in\mathcal{S},
\end{equation}
where $\sigma>0$ is the standard deviation defined according to the confidence of the CSI estimation.
A higher confidence corresponds to a smaller $\sigma$ to concentrate the significance $q^{(\boldsymbol{\Delta})}$ on the original estimate.
With the explicitly defined $\{q^{(\boldsymbol{\Delta})}\}$, Lemma \ref{Lemma 1} completes the robust AHAD criterion design via \eqref{eq: robust AHAD}, whose implementation will be discussed next.

\subsection{Implementation and Discussion}\label{subsec: Implementation}
To implement the robust AHAD criterion \eqref{eq: robust AHAD}, we introduce an auxiliary variable $\mathbf{A}\in\mathbb{R}^{H\times W\times C}$ and employ the elementwise sigmoid function $s: \mathbb{R}^{H\times W\times C}\rightarrow [0,1]^{H\times W\times C}$ \cite{sigmoidpytorch}.
Subsequently, we define the relation between the auxiliary variable and the perturbed HSI as $\widetilde{\mathbf{Y}}_A^{(\boldsymbol{\Delta})}\triangleq s\bigl(\mathcal{T}_{\boldsymbol{\Delta}}(\widetilde{\mathbf{Y}}_H)
+\mathbf{A}\bigr)$.
Furthermore, since $\widetilde{\mathbf{Y}}_A^{(\boldsymbol{\Delta})}=\mathcal{T}_{\boldsymbol{\Delta}}(\widetilde{\mathbf{Y}}_H)+\mathbf{P}$, the anti-detection signal $\mathbf{P}$ can be consequently characterized in terms of the auxiliary variable $\mathbf{A}$, i.e.,
\begin{align}\label{eq: new P}
\mathbf{P}
\equiv 
s\bigl(\mathcal{T}_{\boldsymbol{\Delta}}(\widetilde{\mathbf{Y}}_H)
+\mathbf{A}\bigr)-\mathcal{T}_{\boldsymbol{\Delta}}(\widetilde{\mathbf{Y}}_H).
\end{align}
As the sigmoid function guarantees the practically feasible domain for the perturbed HSI, we reformulate the box-constrained optimization problem \eqref{eq: robust AHAD} w.r.t. $\mathbf{P}$ as an unconstrained counterpart w.r.t. $\mathbf{A}$ based on the relation \eqref{eq: new P}, i.e.,
\begin{align}\label{eq: soften robust AHAD}
\min_{\mathbf{A}}~
&\sum_{\boldsymbol{\Delta}\in\mathcal{S}}q^{(\boldsymbol{\Delta})}
\mathcal{L}\bigl[
s(\mathcal{T}_{\boldsymbol{\Delta}}(\widetilde{\mathbf{Y}}_H)+\mathbf{A})
-\mathcal{T}_{\boldsymbol{\Delta}}(\widetilde{\mathbf{Y}}_H)
\,\bigm|\,
\mathcal{T}_{\boldsymbol{\Delta}}(\widetilde{\mathbf{Y}}_H)
\bigr].
\end{align}
After making it unconstrained, the optimization problem \eqref{eq: soften robust AHAD} remains inherently non-convex and challenging.
In this article, we employ the adaptive moment estimation (Adam) algorithm \cite{Adam} for nonconvex optimization to optimize \eqref{eq: soften robust AHAD} w.r.t. $\mathbf{A}$.
Consequently, the obtained optimal solution $\mathbf{A}^{\star}$ and the predefined surrogate HSI $\widetilde{\mathbf{Y}}_H$ are used to estimate the desired robust anti-detection signal $\mathbf{P}^{\star}$, i.e.,
%
\begin{align}\label{eq: unified signal}
\mathbf{P}^{\star}
&=
\sum_{\boldsymbol{\Delta}\in\mathcal{S}} 
q^{(\boldsymbol{\Delta})}
\bigl[
s\bigl(\mathcal{T}_{\boldsymbol{\Delta}}(\widetilde{\mathbf{Y}}_H)
+\mathbf{A}^{\star}\bigr)-\mathcal{T}_{\boldsymbol{\Delta}}(\widetilde{\mathbf{Y}}_H)
\bigr].
\end{align}

Overall, the proposed criterion effectively incorporates the key characteristics (C1)-(C4) of the challenging AHAD task.
The energy term $\|\mathbf{P}\|_F$ and box-constraint [{see} \eqref{eq: attack_overall}] ensure the EE property in (C3), the SSTV regularization [{see} \eqref{eq: SSTV}] enhances the RR requirement in (C4), and the ARAB regularization [{see} \eqref{eq: REG2}] strongly promotes the ARAB property in (C2).
Subsequent experiments will further show that $\mathbf{P}^{\star}$ can indeed substantially degrade the performance of most mainstream HAD algorithms under different pixel-shifting cases, thereby addressing the black-box detector uncertainty in (C1) and the uncertainty of CSI ({see} Section \ref{subsec: perfect CSI Formulation}).
In contrast, existing perturbation schemes ({see} {Fig.}~\ref{fig: intro_fig}) inevitably violate at least one of the four characteristics (C1)-(C4).
Specifically, the noise effect fails to meet the RR requirement, whereas neither the HAA nor the minus anomaly signal satisfies the ARAB property ({see} Section \ref{sec: Related Work}), rendering existing perturbation schemes less reliable.

\section{Experiments}\label{sec: Experiment}
The section organization is provided as follows.
Initially, Section \ref{subsec: exp setting} describes the experimental settings, including details of the considered security-critical datasets, the experimental protocol for synthesizing imperfect CSI, and the parameter settings for both the HAD baselines and the proposed AHAD framework.
In particular, ten representative and latest HAD methods are incorporated into the evaluations to ensure diversity, thereby providing experimental evidence for addressing the black-box uncertainty in (C1).

Section \ref{subsec: main exp} then systematically establishes the first evaluation benchmark to assess this pioneering AHAD task. 
Specifically, the performance of HAD methods is commonly evaluated by the area under the curve (AUC) w.r.t. the probability of false alarm (PF) and the probability of detection (PD).
In a typical HAD setting, a higher AUC score generally indicates better detection performance. 
However, AHAD pursues an opposite goal, namely ARAB [{see} (C2)]; thus, the resulting anti-detection signal is expected to degrade the performance of HAD methods.
Since this paper represents the first AHAD study, no existing evaluation benchmark is available to clearly determine how low the AUC score should be to demonstrate the effectiveness of an AHAD solution.
Accordingly, Section \ref{subsec: main exp} first carefully evaluates the AUC scores yielded under the unperturbed and AHAD-perturbed scenarios, which are denoted as $\text{AUC}_{up}$ and $\text{AUC}_{ap}$, respectively.
Subsequently, the new AHAD metric, termed area missing caused by anti-detection (ArmCBA)\footnote{The demonstration is available at: \url{https://github.com/IHCLab/AHAD}.}, is then defined as $(1-\frac{\text{AUC}_{ap}}{\text{AUC}_{up}})\times 100\%~~\!(\uparrow)$\footnote{A reliable anti-detection solution is expected to substantially reduce $\text{AUC}_{ap}$ relative to $\text{AUC}_{up}$, yielding a smaller ratio of $\frac{\text{AUC}_{ap}}{\text{AUC}_{up}}$. Hence, by the definition of the ArmCBA metric, a larger ArmCBA score indicates a better AHAD performance.} to validate the effectiveness of an AHAD solution.
This metric represents the relative performance degradation induced by the anti-detection signal; based on the analyses of the qualitative and quantitative performance ({see} Section \ref{subsubsec: Evaluation under Perfect CSI}), ArmCBA values of around 10\%, 15\%, and 20\% are considered the reference levels of acceptable, promising, and superior AHAD performance, respectively.

Based on the new metric and evaluation benchmark for AHAD, we can assess the effectiveness of $\mathbf{P}^{\star}$ under both perfect and imperfect CSI conditions, thereby demonstrating its robustness to CSI uncertainty.
Section \ref{subsec: Ablation} further provides a detailed analysis of how each regularizer contributes to the effectiveness of AHAD based on the evaluation benchmark.
Section \ref{subsec: Resistance} experimentally evaluates whether the resulting anti-detection signal satisfies the RR property.
Section \ref{subsec: Discussions} presents comprehensive analyses and discussions, such as parameter sensitivity analysis, the effect of surrogate-HSI quality, and baseline perturbation comparisons.
A detailed section organization is additionally provided at the beginning of Section \ref{subsec: Discussions} for better clarity.
Finally, considering the rapid development of RS foundation models, Section \ref{subsubsec: HyperSIGMA} evaluates the effectiveness of the proposed AHAD framework against a foundation model-based detector.
%
\begin{table}[t]
\centering
\caption{Detailed information on the five security-critical real-world datasets.
{\blue For each HAD dataset, $H$ denotes the original spatial size of $H\times H$ pixels.
Moreover, $C$ and $A$ represent the number of spectral bands and the number of annotated anomaly pixels, respectively.
Fig. \ref{fig: dataset} shows the false-color compositions for the zero-shifting case.}
More detailed settings are provided in Section \ref{subsec: exp setting}.}\label{tab: data description}
\begin{tabular}{cccccc} \hline\hline
Datasets  & Location  & $H$  & $C$ & Resolution  & $A$ 
\\ \hline
Airport \uppercase\expandafter{\romannumeral 1} & Los Angeles & $100$    & 205 & 7.1 m  & 87
\\
Airport \uppercase\expandafter{\romannumeral 2} & Los Angeles & $100$     & 205 & 17.2 m & 170
\\
MUUFL   & Gulf Park Campus &   $150$   & 64  & 1.0 m  & 259
\\
Urban \uppercase\expandafter{\romannumeral 1} & Texas Coast &   $100$   & 204  & 7.1 m  & 67
\\
Urban \uppercase\expandafter{\romannumeral 2}& San Diego &   $100$   & 205  & 3.4 m  & 272
\\
\hline\hline
\end{tabular}
\end{table}

\begin{figure}[t]
    \centering
    \includegraphics[width=1\linewidth]{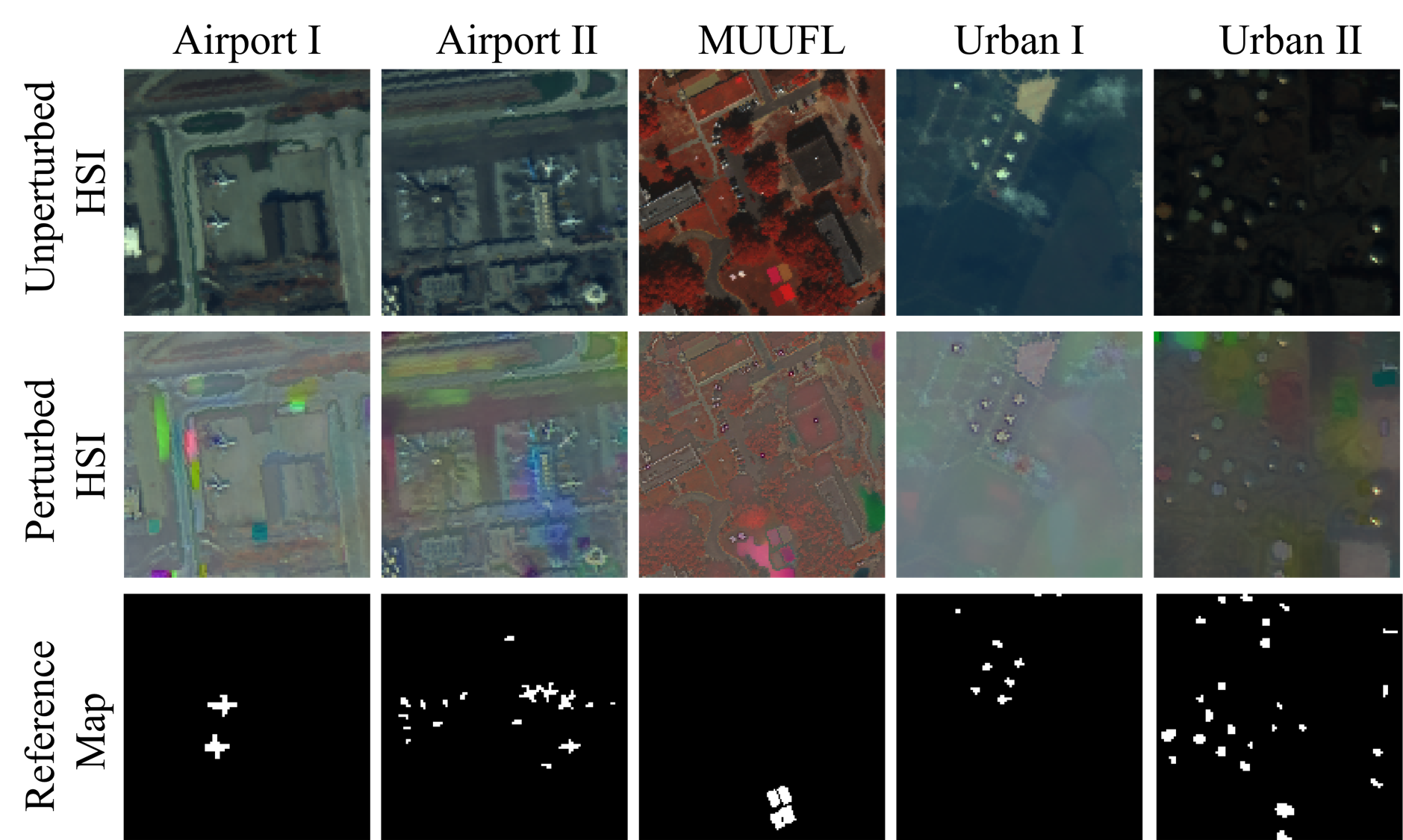}
    \caption{False-color compositions (bands 37, 18, 8 as RGB) of the unperturbed HSIs and AHAD-perturbed HSIs, together with the corresponding reference maps for the five security-critical real-world scenarios under the zero-shifting case. Additional data information can be found in Table \ref{tab: data description}.}\label{fig: dataset}
    \vspace{-0.3cm}
\end{figure}


\begin{figure}[t]
    \centering
    \includegraphics[width=1\linewidth]{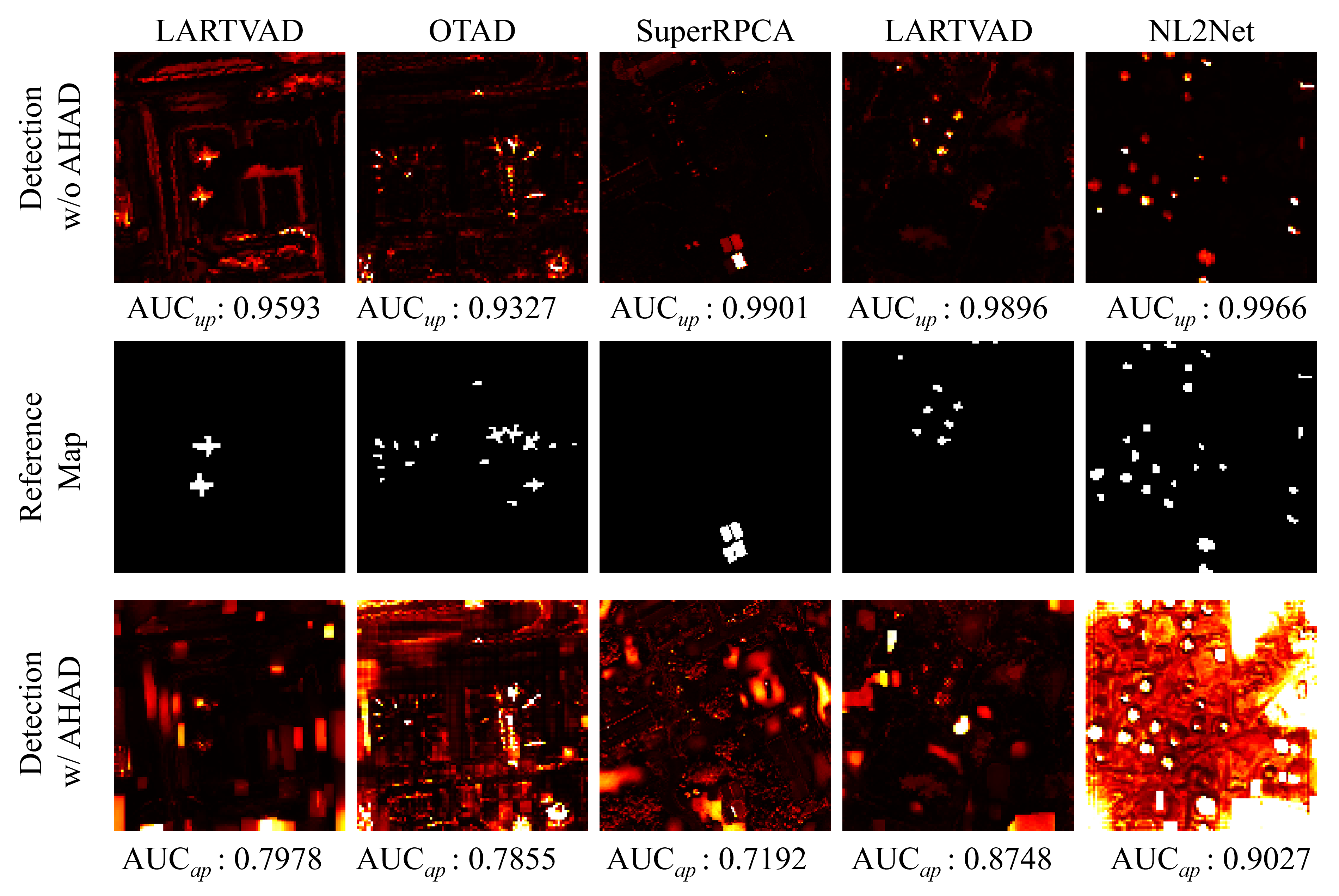}
    \caption{Qualitative comparisons of detection maps under the perfect CSI condition.
    The $\text{AUC}_{up}$ and $\text{AUC}_{ap}$ denote the AUC scores obtained with unperturbed and AHAD-perturbed HSIs as inputs, respectively.
The results show that the proposed AHAD framework can prevent the target area from exposure under the mainstream HAD methods.
 More discussions are presented in Section \ref{subsec: main exp}.
}\label{fig: qualitative}
    \vspace{-0.3cm}
\end{figure}

\subsection{Experimental Settings}\label{subsec: exp setting}
In the subsequent experiments, five widely investigated real-world HAD datasets are incorporated.
These datasets cover airport, campus, and civilian building areas that represent security-sensitive facilities, human activities, and infrastructures.
Such RS scenarios are commonly regarded as security-critical and hence highly related to the AHAD.
The corresponding false-color compositions of the unperturbed and AHAD-perturbed HSIs, together with the reference detection maps, are shown in {Fig.}~\ref{fig: dataset}.
Although the anti-detection signal may not be completely imperceptible at this stage, our primary goal is to prevent detectors from providing reconnaissance information, while meeting (C1)-(C4).
Moreover, Section \ref{subsec: Resistance} will further show that when the detectors are aware of the anti-detection behavior, and attempt to eliminate the signal via some benchmark restoration
technique, these mainstream HAD detectors may undergo even larger performance drops.
From this perspective, the imperceptibility of anti-detection signals may not be necessary, while we still regard this as an important future research line.

Among these datasets, the MUUFL dataset was acquired by Compact Airborne Spectrographic Imager (CASI) sensor \cite{gader2013muufl}, while the remaining datasets were collected by Airborne Visible/Infrared Imaging Spectrometer (AVIRIS)
sensor \cite{AVIRISdata}.
This dataset configuration is consistent with our discussions in Section \ref{sec: Intro}, i.e., the hardware configurations can be reasonably assumed to be a subset of high-standard airborne systems.
Configurations of these datasets, including the location, resolution, and spectral bands, are summarized in Table \ref{tab: data description}.
{\blue To simulate the pixel-shifting effect described in \eqref{eq: shifting}, we first reserve $r$ pixels along each edge of the HAD benchmark datasets, whose original spatial dimensions are provided in Table \ref{tab: data description}.
Then, we regard the remaining area as the surrogate HSI $\widetilde{\mathbf{Y}}_H$.}
For instance, given an HAD dataset $\mathbf{Y}\in \mathbb{R}^{H^{'}\times W^{'}\times C}$ and the maximum pixel-shifting range $r$, the surrogate HSI can be obtained by
\[
\widetilde{\mathbf{Y}}_H:=[\mathbf{Y}]_{(r+1):(H^{'}-r),(r+1):(W^{'}-r),:}.
\]
Accordingly, the reserved pixels in the original HSI can be used to fulfill the boundary pixels of the pixel-shifted HSI $\mathcal{T}_{\boldsymbol{\Delta}}(\widetilde{\mathbf{Y}}_H), \forall \boldsymbol{\Delta}\in\mathcal{S}$.
The sub-reference maps are cropped with the same strategy for all $\boldsymbol{\Delta}$ to ensure fair evaluations.
In this article, $r:=1$ is used to represent a challenging spatial misregistration scenario.
Specifically, a registration error of 0.5 pixels or greater would significantly undermine the reliability of real-world RS tasks \cite{8897135}.
In typical cross-sensor RS applications, even a single-pixel shift can lead to substantial performance degradation \cite{QRCODE}.
These suggest that $r=1$ represents a challenging setting for imperfect CSI.

To assess whether the resulting anti-detection signal satisfies the black-box uncertainty in (C1), we select the HAD baselines based on both effectiveness and diversity.  
Specifically, the reconnaissance detector is assumed to be unknown yet highly effective; the evaluations should cover a wide range of representative and SOTA detectors.
To this end, SuperRPCA \cite{SuperRPCA} is employed as the RPCA-based method, while LRASR \cite{LRASR}, GTVLRR \cite{GTVLRR}, and  LARTVAD \cite{LARTVAD} are adopted as the matrix/tensor-LRR-based approaches.
RGAE \cite{RGAE} is incorporated as a robust benchmark to verify whether the resulting anti-detection signal remains effective against a detector robust to typical noise/outlier corruptions.  
BockNet \cite{BockNet}, PUUNet \cite{PUNNet}, and NL2Net \cite{NL2Net} are included as the SOTA blind-spot HAD networks.
Finally, the latest transformer-based HAD methods, including GTHAD \cite{GT-HAD} and OTAD \cite{OT-AD}, are also included to further strengthen the evaluations.

For the DL-based methods, the default configurations are generally adopted.
The parameter settings of the remaining HAD baselines were partially adopted from the paper or official implementations, with slight dataset-specific adjustments to improve their effectiveness.
For LRASR, the trade-off parameters $\beta$ and $\lambda$ are set to 0.01 for the Urban \uppercase\expandafter{\romannumeral 2} dataset, while 0.01 and 0.05 are adopted for the remaining datasets.
For GTVLRR, the parameters $\beta$ and $\gamma$ are set to 0.5 and 0.01, respectively, across all datasets, while $\lambda$ is selected from \{$2.5\times 10^{-4}$, $1\times 10^{-3}$, $2\times 10^{-2}$\}.
For LARTVAD, the parameters ($K$, $\beta$, $\lambda$) are fixed at (80, 3, 1) for the Airport \uppercase\expandafter{\romannumeral 1} dataset, while (80, 3, 6) is adopted for the other scenarios.  
For SuperRPCA, the parameters ($N$, $t$, $S$, $m$ $\lambda_a$, $\lambda_b$) are slightly adjusted based on the settings in the official implementation.
The settings of each HAD method are consistently used for both unperturbed and AHAD-perturbed conditions to ensure fair evaluations.

For the proposed AHAD framework, the scaling factor of D2CM is set to $p({8\times 10^{2}})$, where $p:=\max(\widetilde{\mathbf{Z}}_{H})^{-2}$, while the standard deviation is fixed as $\sigma:=1$ across all datasets.
Let $(\alpha^{'},\beta^{'},\gamma^{'}):=(2,1,2)$; the trade-off parameters of the SSTV regularizer can be represented as $s(\alpha^{'},\beta^{'},\gamma^{'})$.
For Airport \uppercase\expandafter{\romannumeral 1} and Urban \uppercase\expandafter{\romannumeral 2} datasets, $s$ is set to $8\times 10^{-4}$, whereas $s:=1\times 10^{-3}$ is used for the remaining datasets.
The parameter $\lambda_1$ is set to $2.5\times 10^{-1}$ for Airport \uppercase\expandafter{\romannumeral 1} and Urban \uppercase\expandafter{\romannumeral 1} datasets, and $1\times 10^{-1}$ for the remaining datasets.
Besides, $\lambda_2$ is set to $2\times 10^{-1}$ and $1\times 10^{-1}$ for MUUFL and Urban \uppercase\expandafter{\romannumeral 1} dataset, respectively, while $\lambda_2:=5\times 10^{-1}$ is adopted for the remaining datasets. 
Based on the above empirical settings, we suggest the default configurations of $s:=1\times 10^{-3}$, $\lambda_1:=1\times 10^{-1}$, and $\lambda_2:=5\times 10^{-1}$.
Although the parameters require data-specific tuning, their adjustments remain minor relative to the default settings.
Therefore, we consider adaptive parameter selection and unification an important direction for future work.
The learning rate of 0.05, the number of iterations of 500, and all-zero initialization are applied across all scenarios, resulting in sub-minute computation time.

For the experimental environments, SuperRPCA is performed on a laptop with a 2.50-GHz i7-11700 CPU and 24 GB of RAM.
The remaining experiments are conducted on an Ubuntu 22.04.5 LTS-powered server equipped with a GeForce RTX 3090 and a Ryzen 9 5950X 16-core processor with 128 GB of memory.
The implementation platforms are  MATLAB R2023a, Python 3.10.9, and Torch 2.0.0.
\subsection{Qualitative, Quantitative, and Robustness Evaluations}\label{subsec: main exp}
\subsubsection{Evaluation Benchmark for AHAD}\label{subsubsec: benchmarking}
Since this paper presents the first AHAD study, no generally accepted evaluation benchmark is available, necessitating its establishment.
In the AHAD, a reliable anti-detection signal is expected to result in substantial performance (including qualitative and quantitative) degradation for HAD methods.
In the RS field, the quantitative performance is commonly evaluated using the AUC(PD, PF) score \cite{TGFA-AD,HyFuHAD}.
For conciseness, we denote the AUC scores obtained under unperturbed and AHAD-perturbed conditions as $\text{AUC}_{up}$ and $\text{AUC}_{ap}$, respectively. 
Together with the corresponding qualitative performance, we will establish a criterion for determining whether the observed degradation is sufficient to evidence the effectiveness of the AHAD solution, as detailed below.
The following benchmark development is based on the zero-shifting scenario to avoid potential bias from imperfect CSIs.
In real-world applications, a reconnaissance detector may correspond to unknown yet effective HAD methods [{see} (C1)].
Building on this, the strongest HAD methods for each dataset, measured by $\text{AUC}_{up}$ ({see} Table \ref{tab: quantitative}), are selected as representative examples to establish the benchmarking.
Moreover, we note that some detection maps may achieve promising AUC scores while suffering from intensity over-suppression \cite{TGFA-AD}.
In this context, they can still provide meaningful reconnaissance information, while the qualitative performance remains limited.
To ensure consistency between qualitative performance and the AUC scores, the detection maps are visually adjusted using a capped-square transform \cite{SuperRPCA}, with the threshold selected from \{$5\times10^{-2}$, $2\times10^{-1}$, $8\times10^{-1}$\}.
%
%
The benchmarking discussion is presented as follows.
%
%

%
The detection maps and AUC scores are shown in {Fig.}~\ref{fig: qualitative}.
For the two airport datasets, LARTVAD and OTAD achieve the $\text{AUC}_{up}$ of 0.9593 and 0.9327, respectively.
At these AUC levels, the detection results are expected to highlight the true anomalies (i.e., aircraft) and show slight false detections from the background.
Nevertheless, when the customized anti-detection signals are applied, their $\text{AUC}_{ap}$ falls below 0.8.
At such low AUC levels, false detections become considerably more severe, whereas true anomalies become difficult to distinguish, as shown in the first and second columns of {Fig.}~\ref{fig: qualitative}.
These observations suggest that detection maps with $\text{AUC}_{ap}$ scores around or below 0.8 provide limited information for RS detection tasks. 

Subsequently, SuperRPCA and LARTVAD yield $\text{AUC}_{up}$ of approximately 0.99 on the MUUFL and Urban \uppercase\expandafter{\romannumeral 1} datasets, respectively.
At such effective AUC levels, the true anomalies are well identified in the detection maps, while only very limited background components remain observable.   
In contrast, their detection performance degraded from 0.99 to around 0.7 and from 0.989 to 0.87, respectively, after applying the anti-detection signal.
These observations indicate that even superior detectors can be significantly affected by a customized anti-detection signal.
The detection maps with an $\text{AUC}_{ap}$ of 0.7 further support our conclusion that an $\text{AUC}_{ap}$ of 0.8 or less generally fails to provide meaningful information.
For the $\text{AUC}_{ap}$ from 0.85 to 0.9, the corresponding detection maps still contain pronounced false detections, whereas the true anomalies can only be weakly distinguishable from them.
Finally, for $\text{AUC}_{up}$ above 0.995, as shown in the NL2Net detection map in {Fig.}~\ref{fig: qualitative}, the results are expected to closely match the reference maps.
The above observations naturally lead to the following remark:
\begin{Remark}\label{remark 1}
When $\text{AUC}_{ap}$ is around or below 0.8, the anti-detection signal achieves a reliable protection for the target area.
When $\text{AUC}_{ap}$ ranges from 0.85 to approximately 0.9, the detection map still suffers from noticeable false detections, and the real anomalies can only be weakly identified, suggesting an acceptable anti-detection performance.
\hfill$\square$
\end{Remark}
On the other hand, observations under the unperturbed condition are also summarized to facilitate subsequent analysis.
For $\text{AUC}_{up}$ ranging from 0.93 to approximately 0.95, the detection maps become more informative.
The real anomalies can be readily distinguished, though false-detection suppression still requires further improvement.
The observation indicates that when the $\text{AUC}_{ap}$ can only be reduced to this level, the anti-detection signal remains insufficient, providing only limited protection for the target area.
When the AUC score reaches around or above 0.99, the detection maps appear highly effective and visually consistent with the manually annotated reference map.

\subsubsection{AHAD Evaluation under Perfect CSI}\label{subsubsec: Evaluation under Perfect CSI}
Based on Remark \ref{remark 1}, the effectiveness of the proposed AHAD solution can be fairly evaluated.
We first examine the perfect CSI condition, and summarize the quantitative results in Table \ref{tab: quantitative}.
For the two challenging airport datasets, most HAD methods achieve $\text{AUC}_{up}$ above 0.9.
Nevertheless, after applying the anti-detection signal, none of these detectors can achieve an $\text{AUC}_{ap}$ above 0.8 on the Airport \uppercase\expandafter{\romannumeral 1} dataset.
Such low AUC levels barely provide discriminative information for RS anomaly detection ({see} Remark \ref{remark 1}), substantiating the effectiveness of the proposed AHAD framework.
Moreover, the ArmCBA $(1-\frac{\text{AUC}_{ap}}{\text{AUC}_{up}})\times 100\%~~\!(\uparrow)$ is also provided to more intuitively demonstrate its anti-detection capability, as shown in Table \ref{tab: quantitative}.

%
In the Airport \uppercase\expandafter{\romannumeral 1} dataset, the customized anti-detection signal generally leads to a substantial performance drop of 20\%, and particularly around 30\% for several HAD methods (e.g., LRASR, GTVLRR, and SuperRPCA). 
These observed ArmCBAs indicate that, even when the detectors can achieve an ideal detection under the unperturbed condition ($\text{AUC}_{up}=1$), an ArmCBA of 20\% or above (i.e., $\text{AUC}_{ap}\leq0.8$) makes the resulting detection maps uninformative for the ground facilities ({see} Remark \ref{remark 1}).  
For the Airport \uppercase\expandafter{\romannumeral 2} dataset, all HAD baselines fail to achieve an $\text{AUC}_{ap}$ higher than 0.84 when the anti-detection signal is applied.
The corresponding performance drops also mostly range from 10\% to 20\%, and can exceed 30\% in certain methods (e.g., LRASR).
Even the robust HAD method, RGAE, shows a drop of over 20\%, indicating that the proposed AHAD framework provides reliable anti-detection capability against the RS detection of ground facilities.
In the MUUFL dataset, multiple advanced HAD methods, including LRASR, GTVLRR, LARTVAD, SuperRPCA, RGAE, and NL2Net, yield $\text{AUC}_{up}$ close to or even above 0.99.
These results suggest that their detection maps show promising qualitative performance and can provide highly reliable information for real-world detection tasks ({see} Section \ref{subsubsec: benchmarking}).
However, the non-DL HAD methods are notably sensitive to the customized anti-detection signal.
For instance, the $\text{AUC}_{ap}$ of LRASR, GTVLRR, LARTVAD, and SuperRPCA reduce to 0.7914, 0.7255, 0.7562, and 0.7192, respectively, which corresponds to at least a 19.4\% performance drop.
For advanced DL-based methods, the anti-detection signal can still reduce their $\text{AUC}_{ap}$ to approximately 0.8-0.9, suggesting acceptable anti-detection performance ({see} Remark \ref{remark 1}).
Similar performance degradations can also be observed on the Urban \uppercase\expandafter{\romannumeral 1} and Urban \uppercase\expandafter{\romannumeral 2} dataset. 
Specifically, the top-performing HAD algorithms on these datasets, including LARTVAD, SuperRPCA, RGAE, BockNet, PUUNet, NL2Net, and OTAD, achieve $\text{AUC}_{up}$ ranging from around 0.98 to 0.99.
Nevertheless, the customized anti-detection signal generally results in a 10\% ArmCBA for these superior methods.
Such performance degradation may be considered acceptable, as it reduces the reliability of the corresponding detection maps, thereby providing a certain degree of protection and enhancing safety for the target area.

Although the proposed AHAD framework effectively degrades the detection performance of a wide range of HAD methods, its effectiveness may vary across scenarios, and occasional failures are observed.
A representative example is observed in the GTHAD detection on the MUUFL dataset.
In this case, the AUC slightly increases after applying the AHAD signal, thereby resulting in a negative ArmCBA score.
This failure case is potentially attributed to the global modeling capability of transformer architectures and the specialized dual-branch anomaly/background gating mechanism.
Nevertheless, when evaluated against GTHAD across all considered datasets, the customized AHAD signal still reduces the average AUC score from 0.9347 to 0.8527, approaching the reference level of acceptable performance.
To further improve the generalizability and reliability of the proposed AHAD framework against different HAD methods, several directions are discussed in Section \ref{sec: Conclusion}.

In general, under the unperturbed condition, mainstream HAD baselines can effectively detect anomalies, with an average $\text{AUC}_{up}$ of approximately 0.95 across these security-critical scenarios, as summarized in the last row of Table \ref{tab: quantitative}.
Furthermore, building upon Remark \ref{remark 1}, an $\text{AUC}_{ap}$ of 0.8 and 0.85 correspond to reliable and acceptable anti-detection results, respectively. 
Motivated by these observations, the following remark is provided to systematically characterize the ArmCBA:
\begin{Remark}\label{remark 2}
ArmCBA values of approximately $10.5\%=(1-\frac{0.85}{0.95})$, $15.8\%=(1-\frac{0.80}{0.95})$, and $20\%=(1-\frac{0.8}{1.0})$ {\blue are used in this study as empirical reference levels for indicating} the acceptable, promising, and superior AHAD performance, respectively, under the considered conditions. 
\hfill$\square$
\end{Remark}
As reported in the last row of Table \ref{tab: quantitative}, the resulting anti-detection signal generally induces an average 15\% ArmCBA across real-world scenarios under the zero-shifting case.  
In particular, the ArmCBA can reach close to or exceed 20\% for several detectors, such as LRASR, SuperRPCA, and BockNet, substantiating the effectiveness of the proposed AHAD framework.
Moreover, even if the strongest HAD method (i.e., LARTVAD), which yields an average $\text{AUC}_{up}$ of 0.9622, is adopted, the proposed AHAD framework can significantly reduce its $\text{AUC}_{ap}$ to around 0.8, preventing it from providing reconnaissance information for the ground facilities.

\subsubsection{Robustness Against Imperfect CSI}\label{subsubsec: Robustness}
As evidenced by the above evaluations ({see} Table \ref{tab: quantitative}), the proposed robust AHAD solution ({see} Section \ref{subsec: Robust}) demonstrates high potential for real-world anti-detection protection under the zero-shifting case, thereby supporting the primary goal of its non-robust counterpart. 
This observation is the main reason this paper presents the implementation and evaluation based on the robust AHAD formulation, as it provides a more general solution to the real-world anti-detection task than its non-robust counterpart.
In the following, we further demonstrate that the robust AHAD solution remains consistently effective even under imperfect CSIs.

Based on the observations summarized in Remark \ref{remark 2}, such robustness assessment can be conducted by evaluating the average ArmCBA under various pixel-shifting cases.
Specifically, for each dataset, the proposed framework estimates a unified anti-detection signal $\mathbf{P}^{\star}$ via the robust AHAD criterion [{see} \eqref{eq: unified signal}].
To assess its robustness against imperfect CSI, $\mathbf{P}^{\star}$ is directly applied to all pixel-shifted HSIs, i.e., $\widetilde{\mathbf{Y}}_A^{(\boldsymbol{\Delta})}=\mathcal{T}_{\boldsymbol{\Delta}}(\widetilde{\mathbf{Y}}_H)+\mathbf{P}^\star,~\forall \boldsymbol{\Delta}\in\mathcal{S}$.
In this context, the corresponding average $\text{AUC}_{up}$ and $\text{AUC}_{ap}$ under different pixel-shifting cases are first computed, and the corresponding ArmCBA can be derived accordingly.
A robust AHAD solution is expected to yield a consistently significant drop in AUC under both perfect and imperfect CSI.

\begin{figure*}[t]
    \centering
    \includegraphics[width=1\linewidth]{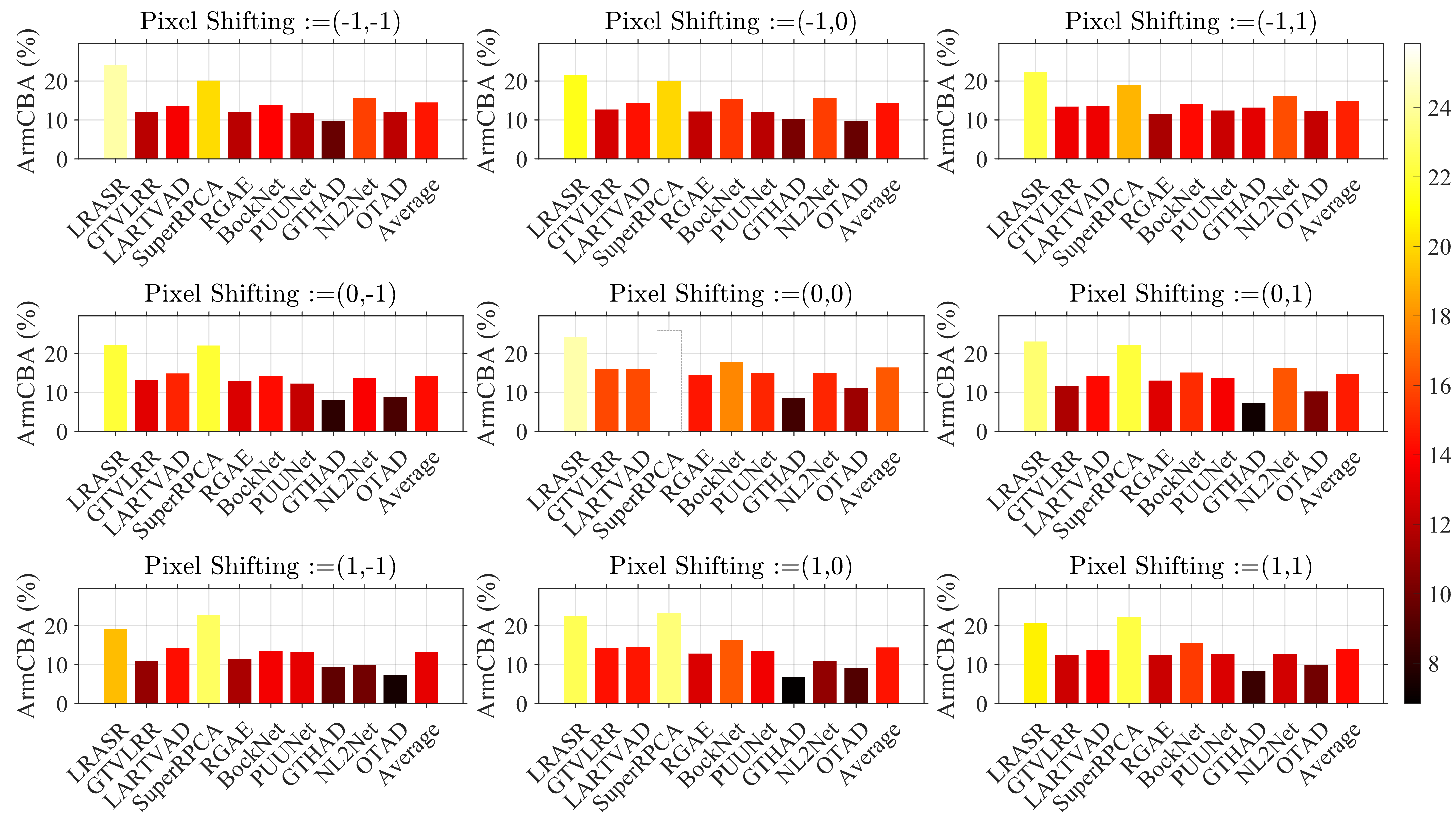}
    \caption{Histogram of the ArmCBA ({see} Table \ref{tab: quantitative}) across all pixel-shifting cases under the setting of $r=1$.
    %
   %
   These results indicate that the robust anti-detection signal $\mathbf{P}^{\star}$ generally leads to an ArmCBA of around 15\% across all pixel-shifting cases for mainstream HAD methods, demonstrating the robustness of our AHAD framework in the imperfect CSI scenario ({see} Remark \ref{remark 2}).
   Further discussions are presented in Section \ref{subsubsec: Robustness}.
    }\label{fig: shifting_drop}
    \vspace{-0.3cm}
\end{figure*}

As illustrated in {Fig.}~\ref{fig: shifting_drop}, the results strongly support the robustness of the proposed AHAD solution against all pixel-shifting scenarios under the settings of $r=1$.
Specifically, the resulting anti-detection signals $\mathbf{P}^{\star}$ consistently cause substantial performance degradation of a wide range of mainstream HAD methods.
In general, the ArmCBA reaches approximately 15\%, and possibly more under different pixel-shifting cases ({see} {Fig.}~\ref{fig: shifting_drop}).
Building upon Remark \ref{remark 2}, the above observations suggest an effective AHAD performance regardless of perfect or imperfect CSI.
In a nutshell, based on the above evaluations, the proposed AHAD framework can degrade the performance of a wide range of HAD baselines, including DL-based and non-DL-based methods, across different security-critical real-world scenarios.
Although the effectiveness of the proposed framework may vary across conditions at this early stage of development, the current results have shown considerable robustness against imperfect CSI and highlight its strong potential for achieving practical AHAD application.

\begin{table*}[t]
\footnotesize
\centering
\caption{Quantitative comparisons, reported in AUC(PD, PF)$~\!(\uparrow)$, on the five representative real-world datasets under perfect CSI. 
To clearly distinguish the AUC scores obtained under the unperturbed and AHAD-perturbed conditions, we denote them by $\text{AUC}_{up}$ and $\text{AUC}_{ap}$, respectively.
The ArmCBA metric is then defined as $(1-\frac{\text{AUC}_{ap}}{\text{AUC}_{up}})\times 100\%~~\!(\uparrow)$, which represents the relative performance degradation. 
{\blue As demonstrated, the customized anti-detection signal yields average ArmCBA scores above 14.5\% for eight evaluated HAD methods.
Among these methods, three detectors further achieve mean ArmCBA scores between 15\% and 20\%, and two methods are even higher than 20\%}, indicating a meaningful degree of AHAD performance ({see} Remark \ref{remark 2}) and considerable potential for future development.
Further analytical discussion is provided in Section \ref{subsec: main exp}.
}\label{tab: quantitative}

\begin{tabular}{cc||cccccccccc}
\hline \hline
Scenario & Metric 
& LRASR & GTVLRR & LARTVAD & SuperRPCA & RGAE & BockNet & PUUNet & GTHAD & NL2Net & OTAD   \\

\hline
\rowcolor{blue!8}
& $\text{AUC}_{up}$
& 0.9012 & 0.8962 & 0.9593 & 0.8559 & 0.7465 & 0.9161 & 0.9204 & 0.9472 & 0.7042 & 0.9347   \\

\rowcolor{blue!8}
& $\text{AUC}_{ap}$
& 0.5797 & 0.5712 & 0.7978 & 0.6127 & 0.6663 & 0.7151 & 0.7226 & 0.7548 & 0.5573 & 0.7403 \\

\rowcolor{red!8}
\multirow{-3}{*}{Airport \uppercase\expandafter{\romannumeral 1}}
& ArmCBA (\%) 
& 35.6747 & 36.2642 & 16.8352 & 28.4145 & 10.7435 & 21.9408 & 21.4907 & 20.3125 & 20.8606 & 20.7981  \\

\hline
\rowcolor{blue!8}
& $\text{AUC}_{up}$ 
& 0.8609 & 0.9038 & 0.8960 & 0.9265 & 0.8881 & 0.9211 & 0.9230 & 0.9259 & 0.9240 & 0.9327    \\

\rowcolor{blue!8}
& $\text{AUC}_{ap}$
& 0.6021 & 0.8387 & 0.7425 & 0.6237 & 0.6919 & 0.7184 & 0.8098 & 0.8293 & 0.7847 & 0.7855  \\

\rowcolor{red!8}
\multirow{-3}{*}{Airport \uppercase\expandafter{\romannumeral 2}}
& ArmCBA (\%) 
& 30.0616 & 7.2029 & 17.1317 & 32.6821 & 22.0921 & 22.0063 & 12.2644 & 10.4331 & 15.0758 & 15.7821  \\

\hline
\rowcolor{blue!8}
& $\text{AUC}_{up}$ 
& 0.9819 & 0.9901 & 0.9728 & 0.9901 & 0.9842 & 0.9377 & 0.9623 & 0.8459 & 0.9875 & 0.9161   \\

\rowcolor{blue!8}
& $\text{AUC}_{ap}$ 
& 0.7914 & 0.7255 & 0.7562 & 0.7192 & 0.9061 & 0.6879 & 0.8003 & 0.8561 & 0.9245 & 0.8851  \\

\rowcolor{red!8}
\multirow{-3}{*}{MUUFL}
& ArmCBA (\%) 
& 19.4012 & 26.7246 & 22.2656 & 27.3609 & 7.9354 & 26.6397 & 16.8347 & -1.2058 & 6.3797 & 3.3839  \\

\hline
\rowcolor{blue!8}
& $\text{AUC}_{up}$
& 0.8708 & 0.9430 & 0.9896 & 0.9372 & 0.9818 & 0.9792 & 0.9859 & 0.9772 & 0.9837 & 0.9896   \\

\rowcolor{blue!8}
& $\text{AUC}_{ap}$
& 0.8171 & 0.9389 & 0.8748 & 0.8135 & 0.8468 & 0.8480 & 0.8593 & 0.9297 & 0.7568 & 0.9090 \\

\rowcolor{red!8}
\multirow{-3}{*}{Urban \uppercase\expandafter{\romannumeral 1}}
& ArmCBA (\%) 
& 6.1667 & 0.4348 & 11.6006 & 13.1989 & 13.7503 & 13.3987 & 12.8411 & 4.8608 & 23.066 & 8.1447   \\


\hline
\rowcolor{blue!8}
& $\text{AUC}_{up}$ 
& 0.9681 & 0.9713 & 0.9931 & 0.9919 & 0.9947 & 0.9683 & 0.9805 & 0.9773 & 0.9966 & 0.9888   \\

\rowcolor{blue!8}
& $\text{AUC}_{ap}$ 
& 0.6763 & 0.8845 & 0.8737 & 0.7187 & 0.8169 & 0.9224 & 0.8699 & 0.8938 & 0.9027 & 0.9134  \\

\rowcolor{red!8}
\multirow{-3}{*}{Urban \uppercase\expandafter{\romannumeral 2}}
& ArmCBA (\%) 
& 30.1415 & 8.9365 & 12.0230 & 27.5431 & 17.8747 & 4.7403 & 11.2800 & 8.5439 & 9.4220 & 7.6254  \\

\hline
\rowcolor{blue!8}
& $\text{AUC}_{up}$ 
& 0.9166 & 0.9409 & 0.9622 & 0.9403 & 0.9191 & 0.9445 & 0.9544 & 0.9347 & 0.9192 & 0.9524     \\

\rowcolor{blue!8}
& $\text{AUC}_{ap}$ 
& 0.6933 & 0.7918 & 0.8090 & 0.6976 & 0.7856 & 0.7784 & 0.8124 & 0.8527 & 0.7852 & 0.8467   \\

\rowcolor{red!8}
\multirow{-3}{*}{Average}
& ArmCBA (\%) 
& 24.3618 & 15.8465 & 15.9218 & 25.8109 & 14.5251 & 17.5860 & 14.8785 & 8.7729 & 14.5779 & 11.0983  \\

\hline \hline    
\end{tabular}
\end{table*}
\begin{table*}[t]
\footnotesize
\centering
\caption{Experimental evaluations of the proposed AHAD framework under different ablation settings.
In each setting, the ArmCBA $\!(\uparrow)$ averaged across HAD baselines is adopted as the quantitative metric.
The SSTV, Lipschitz, and PAG columns show whether the corresponding regularizers (REGs) are used, while the Denoising (Pre) column indicates whether the widely used preprocessing (Pre) step is applied before HAD.
Further details on the ablation studies and RR evaluations are provided in Section \ref{subsec: Ablation} and Section \ref{subsec: Resistance}, respectively.
}\label{tab: ablation}
\begin{tabular}{cccc!{\color{black!35}\vrule width 0.6pt}c||cccccc} \hline \hline
               Ablation Case & SSTV \eqref{eq: SSTV} & Lipschitz \eqref{eq: topology-enhanced ARAB} & PAG \eqref{eq: pseudo-anomalies} & Denoising (Pre)& Airport \uppercase\expandafter{\romannumeral 1}  & Airport \uppercase\expandafter{\romannumeral 2}     & MUUFL  & Urban \uppercase\expandafter{\romannumeral 1}    & Urban \uppercase\expandafter{\romannumeral 2}   & Average
                \\   
                 \hline  \rowcolor{red!8}
                Case 1 & & \checkmark &  \checkmark & & 39.2840 & 66.3909 & 30.6482 & 21.7317 & 60.1326 & 43.5439 
                 \\   
                \hline\hline \rowcolor{blue!8}
               Case 2 & \checkmark  & \checkmark &  & & 13.1159 & 10.9921 & 14.4556 & 8.8919 & 8.9140 & 11.2291 
                 \\   
                \hline\rowcolor{blue!8}
               Case 3 & \checkmark &  & \checkmark& & -1.6933 & 0.0835 & 1.1036 & 1.0127 & 0.9155 & 0.3242 
                \\
                 \hline\rowcolor{blue!8}
                Case 4 & \checkmark & \checkmark &\checkmark &  & 23.5023 & 18.4069 & 15.8466 & 10.8332 & 13.8171 & 16.3211  
                \\
                \hline\rowcolor{blue!8}
                Case 5 &  \checkmark & \checkmark &\checkmark & \checkmark & 32.7830 & 18.8684 & 14.4352 & 53.1407 & 13.0994 & 26.3964  
                \\
\hline \hline    
\end{tabular}
\end{table*}
%
\subsection{Ablation Study}\label{subsec: Ablation}
In the following, we analyze the regularization terms in detail to elucidate their respective contributions to the AHAD task.
Specifically, in the proposed AHAD criterion, the SSTV regularization [{see} \eqref{eq: SSTV}] and ARAB regularization [{see} \eqref{eq: REG2}], are incorporated to yield the reliable and practically feasible anti-detection solution.
The ARAB regularizer comprises two components, including Lipschitz-forcing regularization [{see} \eqref{eq: topology-enhanced ARAB}] and pseudo-anomaly generation (PAG) [{see} \eqref{eq: pseudo-anomalies}]. 
To clearly investigate their respective contributions, we regard them as two individual regularizers in this ablation study. 
These settings lead to the four different ablation cases, including 1) Lipschitz-PAG, 2) SSTV-Lipschitz, 3) SSTV-PAG, and 4) SSTV-Lipschitz-PAG (fully regularized), where the last one corresponds to the proposed framework.

These cases correspond to the first through fourth rows of Table \ref{tab: ablation}, in which the effect of the unused regularizer is excluded by setting its trade-off parameter to zero.
For instance, in the first ablation setting (i.e., Lipschitz-PAG), the SSTV regularizer is removed by setting $(\alpha,\beta,\gamma):=(0,0,0)$, while the remaining parameter settings ({see} Section \ref{subsec: exp setting}) are unchanged to ensure a fair evaluation.
As the proposed framework has demonstrated strong robustness against imperfect CSI ({see} Section \ref{subsubsec: Robustness}), the subsequent analyses focus on the zero-shifting case and report the corresponding ArmCBA averaged across the security-critical datasets and baselines.
For the last row in Table \ref{tab: ablation}, we aim to experimentally demonstrate the RR property [{see} (C4)] of the proposed AHAD framework.
The corresponding evaluations are provided in Section \ref{subsec: Resistance}.

First, as shown in Table \ref{tab: ablation}, removing TV regularization on the anti-detection signal seems to yield the most effective AHAD protection, as evidenced by the largest performance degradation of the HAD baselines.
In other words, even if a superior HAD baseline can achieve an ideal $\text{AUC}_{up}$ of 1, the corresponding $\text{AUC}_{ap}$ can be reduced to below 0.7 once an anti-detection signal is applied.
However, the SSTV regularizer should still be introduced to ensure a practically feasible solution [{see} (C3)], rather than merely pursuing stronger AHAD performance. 
Specifically, as detailed in the paragraph following \eqref{eq: SSTV}, without adopting SSTV, the AHAD-perturbed HSI exhibits severe and noticeable perturbation patterns, as shown in the first row of {Fig.}~\ref{fig: restroation_resistance_quali}.
Such severe perturbations imply an extremely high energy requirement for an anti-detection signal, which may be difficult to deploy in practice. 
Consequently, to facilitate real-world application, we should focus on more practical settings, Case 2 to Case 4.

As presented in the second through fourth rows of Table \ref{tab: ablation}, under a more practical setting (i.e., with SSTV), the Lipschitz-forcing regularizer contributes much more significantly than the PAG regularizer.
When only the Lipschitz-forcing regularizer is used for (C2), the resulting anti-detection signal still induces an ArmCBA of approximately 10\%, or even larger, across mainstream HAD baselines ({see} Case 2 in Table \ref{tab: ablation}).
Based on Remark \ref{remark 2}, this result still indicates acceptable AHAD performance.
Conversely, when only the individual PAG regularizer is used for (C2), the results suggest a marginal or even negative impact on AHAD performance in terms of ArmCBA ({see} Case 3 in Table \ref{tab: ablation}).
This may be because real-world anomalies usually deviate from background components substantially, whereas the generated pseudo-anomaly may introduce insufficiently discriminative anomalous patterns ({see} the third row of {Fig.}~\ref{fig: restroation_resistance_quali}).
Accordingly, the PAG and Lipschitz-forcing regularizers should be adopted jointly, namely the ARAB regularizer [{see} \eqref{eq: REG2}]. 
Once the discrepancy between true anomalies and background components is eliminated by the Lipschitz-forcing regularizer, the induced pseudo-anomalies can effectively fool the detectors.
This joint effect results in a more substantial ArmCBA improvement, as shown in the fourth row of Table \ref{tab: ablation}.
Overall, under a more practically feasible setting with SSTV, the ablation studies suggest that all regularizers in the proposed framework play essential roles in achieving reliable AHAD protection. 

\subsection{Evaluation of Restoration Resistance}\label{subsec: Resistance}
\begin{figure}[t]
    \centering
    \includegraphics[width=1\linewidth]{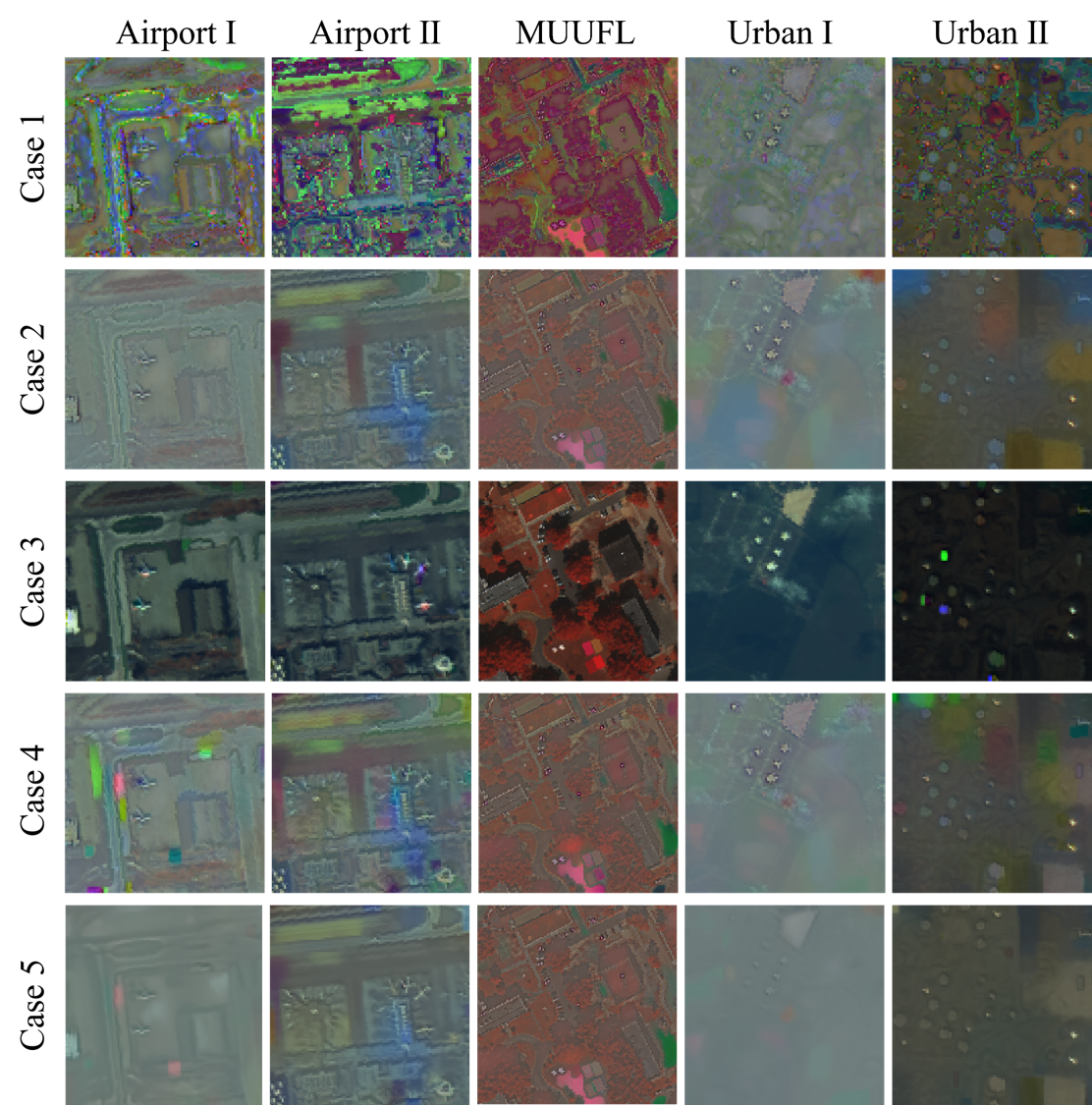}
    \caption{Qualitative comparisons of the AHAD-perturbed HSIs under different ablation settings, referred to as Case 1 to Case 5, whose configurations correspond to the first through fifth rows of Table \ref{tab: ablation}.
    Cases 1 to 3 examine the effects of each regularizer, including SSTV [{see} \eqref{eq: SSTV}], Lipschitz-forcing [{see} \eqref{eq: topology-enhanced ARAB}], and PAG [{see} \eqref{eq: pseudo-anomalies}].
    Case 4 represents the proposed AHAD, which yields the most practical and effective solution ({see} Section \ref{subsec: Ablation}).
    Case 5 presents GLF-restored \cite{GLF} results of Case 4 to investigate the RR property.
    The results indicate that the commonly used denoising technique may further degrade detection performance, as evaluated in Section \ref{subsec: Resistance}.
    }\label{fig: restroation_resistance_quali}
    \vspace{-0.3cm}
\end{figure}
%
This section provides an experimental demonstration of the RR property for the resulting anti-detection signal.
As presented in (C4), reliable anti-detection signals should not be easily removable from the perturbed HSI; otherwise, their reliability and practicality would be substantially undermined.
In the RS field, well-established denoising techniques are commonly used as a preprocessing step \cite{HyperKING,9318503} to enhance the reliability of subsequent applications.
In practice, reconnaissance detectors may also employ such advanced techniques to mitigate underlying noise corruption, thereby ensuring effective detection.
To evaluate the RR property of the resulting anti-detection signal, we first apply a hyperspectral denoising method on the AHAD-perturbed HSI, followed by performing HAD to investigate whether the restoration technique can reduce the ArmCBA score.

In this experiment, the representative global and nonlocal low-rank factorizations (GLF) denoiser \cite{GLF} is used as the restoration technique.
For the GLF denoiser settings, the subspace dimension is fixed at 8, as suggested in the official implementation, while the perturbation type is set to ``additive'' based on the relationship between the anti-detection signal and the AHAD-perturbed HSIs $\widetilde{\mathbf{Y}}_A=\widetilde{\mathbf{Y}}_ H+\mathbf{P}^{\star}$. 
Subsequently, the qualitative comparisons of the AHAD-perturbed HSIs without and with GLF restoration are shown in the fourth and fifth rows of {Fig.}~\ref{fig: restroation_resistance_quali}, respectively.

According to the results, applying this advanced restoration technique not only fails to eliminate the anti-detection effect in perturbed HSIs but may even lead to additional distortions.
Specifically, typical noise effects introduce irregular, high-frequency corruption in observations; therefore, denoising methods are generally designed to suppress this unfavorable non-smoothness, thereby yielding smoother image structures.
However, the Lipschitz-forcing perturbation, produced via ARAB regularization, is designed to suppress local variations in the HSI, thereby assimilating the real anomalies into the background components.
Consequently, applying such an advanced restoration technique on the AHAD-perturbed HSI may produce excessively smoothed image structures ({see} case 5 in {Fig.}~\ref{fig: restroation_resistance_quali}), rather than mitigating the anti-detection effect.
Furthermore, the induced over-smoothness may further increase the ArmCBA scores, as presented in the last two rows of Table \ref{tab: ablation}.
For instance, the ArmCBA metric increases further, from {\blue 23.5\%} to 32.78\% on the Airport \uppercase\expandafter{\romannumeral 1} dataset and from 10\% to 53\% on the Urban \uppercase\expandafter{\romannumeral 1} dataset, after denoising.
Based on these results, this preprocessing step cannot remove the anti-detection signal and may even further degrade the performance of reconnaissance detectors.
\begin{figure}[t]
    \centering
    \includegraphics[width=1\linewidth]{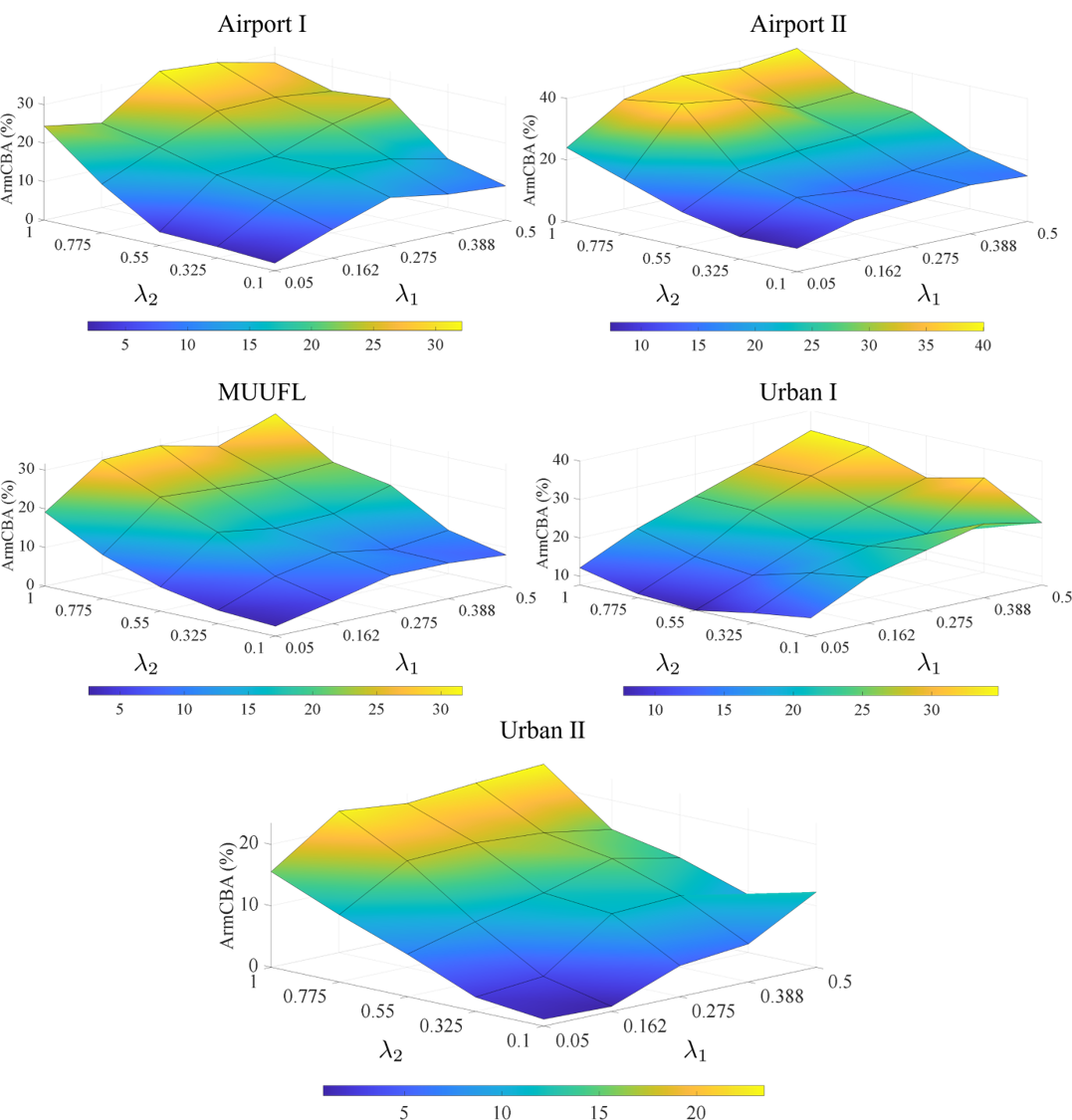}
    \caption{Analysis of parameter sensitivity of the proposed AHAD framework.
    The results suggest that although increasing both $\lambda_1$ [corresponding to Lipschitz REG in \eqref{eq: topology-enhanced ARAB}] and $\lambda_2$ [corresponding to PAG REG in \eqref{eq: pseudo-anomalies}] improves AHAD performance, increasing $\lambda_1$ yields a more substantial enhancement.
    More details are provided in Section \ref{subsubsec: para}.
    }\label{fig: Para_Sen}
    \vspace{-0.3cm}
\end{figure}
%
\subsection{Further Discussion}\label{subsec: Discussions}
In this section, we present additional evaluations and discussions from multiple perspectives to facilitate a deeper understanding.
First, a sensitivity analysis of principal trade-off parameters is presented in Section \ref{subsubsec: para}, providing practical guidelines for future parameter adjustments.
Second, Section \ref{subsubsec: Surrogate error} investigates and discusses the impact of surrogate-HSI quality on AHAD Performance.
Third, Section \ref{subsubsec: p-bnemchmark} presents detailed comparisons between the proposed AHAD framework and several commonly used perturbation strategies, including LRR-based perturbation and noise-based interference.
Finally, Section \ref{subsubsec: larger pixel-shift} further evaluates the robustness of the proposed AHAD framework under more severe pixel-level shifting and sub-pixel misalignments.
Since practical reconnaissance detectors are expected to incorporate strong HAD algorithms, the above analyses focus on the top-performing HAD methods, including LARTVAD, BockNet, PUUNet, and OTAD.

\subsubsection{Sensitivity Analysis of Trade-off Parameters}\label{subsubsec: para}
Although the default parameter settings are recommended in Section \ref{subsec: exp setting}, data-specific adjustments may still be required to facilitate future applications.
Accordingly, we further analyze the parameter sensitivity of the Lipschitz REG and PAG REG to provide guidelines for future users.
Specifically, the proposed AHAD criterion comprises three REGs, namely Lipschitz in \eqref{eq: topology-enhanced ARAB}, PAG in \eqref{eq: pseudo-anomalies}, and SSTV in \eqref{eq: SSTV}. 
The effects of Lipschitz and PAG REGs are controlled by the trade-off parameters of $\lambda_1$ and $\lambda_2$, respectively.
To investigate their individual effect, five values are uniformly sampled over the intervals of $[0.001, 0.1]$ and $[0.1, 1]$ to form the candidate sets for $\lambda_1$ and $\lambda_2$, respectively.
Since the SSTV REG term is introduced to ensure the EE and RR properties of the resulting anti-detection signal, rather than to directly improve AHAD performance, the corresponding trade-off parameter $s$ is fixed at the default value of $1\times 10^{-3}$ throughout this analysis. 
The remaining configurations, including learning rate and number of iterations, are kept unchanged from those reported in Section \ref{subsec: exp setting}. 
With the above settings, the effects of $\lambda_1$ and $\lambda_2$ are evaluated in terms of the ArmCBA metric averaged across all top-performing HAD methods.

The averaged ArmCBA scores are presented for each of the five real-world datasets in {Fig.}~\ref{fig: Para_Sen}.
The results consistently show that increasing either $\lambda_1$ or $\lambda_2$ improves ArmCBA scores in most scenarios, substantiating their positive contributions and necessity.
When $\lambda_1$ is fixed at a relatively small value (e.g., 0.05), the ArmCBA scores are significantly improved as $\lambda_2$ increases in all datasets except Urban \uppercase\expandafter{\romannumeral 1}.  
However, an overly large $\lambda_2$ may introduce more pseudo-anomaly patterns into the resulting AHAD signal, thereby making the perturbations more visually noticeable.
Conversely, when we keep $\lambda_2$ at a relatively small value and increase the value of $\lambda_1$, the ArmCBA scores can still be improved across the five real-world scenarios (more general than increasing $\lambda_2$), while preserving the AHAD-perturbed HSI with a more natural appearance.
Especially, a large $\lambda_1$ tends to introduce smooth, haze-like patterns, rather than visually suspicious perturbations, which is more consistent with the stealthiness goal of this AHAD task.
Overall, simultaneously increasing $\lambda_1$ and $\lambda_2$ results in more substantial improvement than increasing either parameter alone, which further validates their complementary roles.
According to the above analyses and discussion, when a stronger AHAD signal is required, jointly increasing $\lambda_1$ and $\lambda_2$ while prioritizing $\lambda_1$ is recommended to achieve a better improvement and maintain a favorable balance of perturbation stealthiness.

\subsubsection{Effect of Surrogate-HSI Quality on AHAD Performance}\label{subsubsec: Surrogate error}
\begin{figure}[t]
    \centering
    \includegraphics[width=1\linewidth]{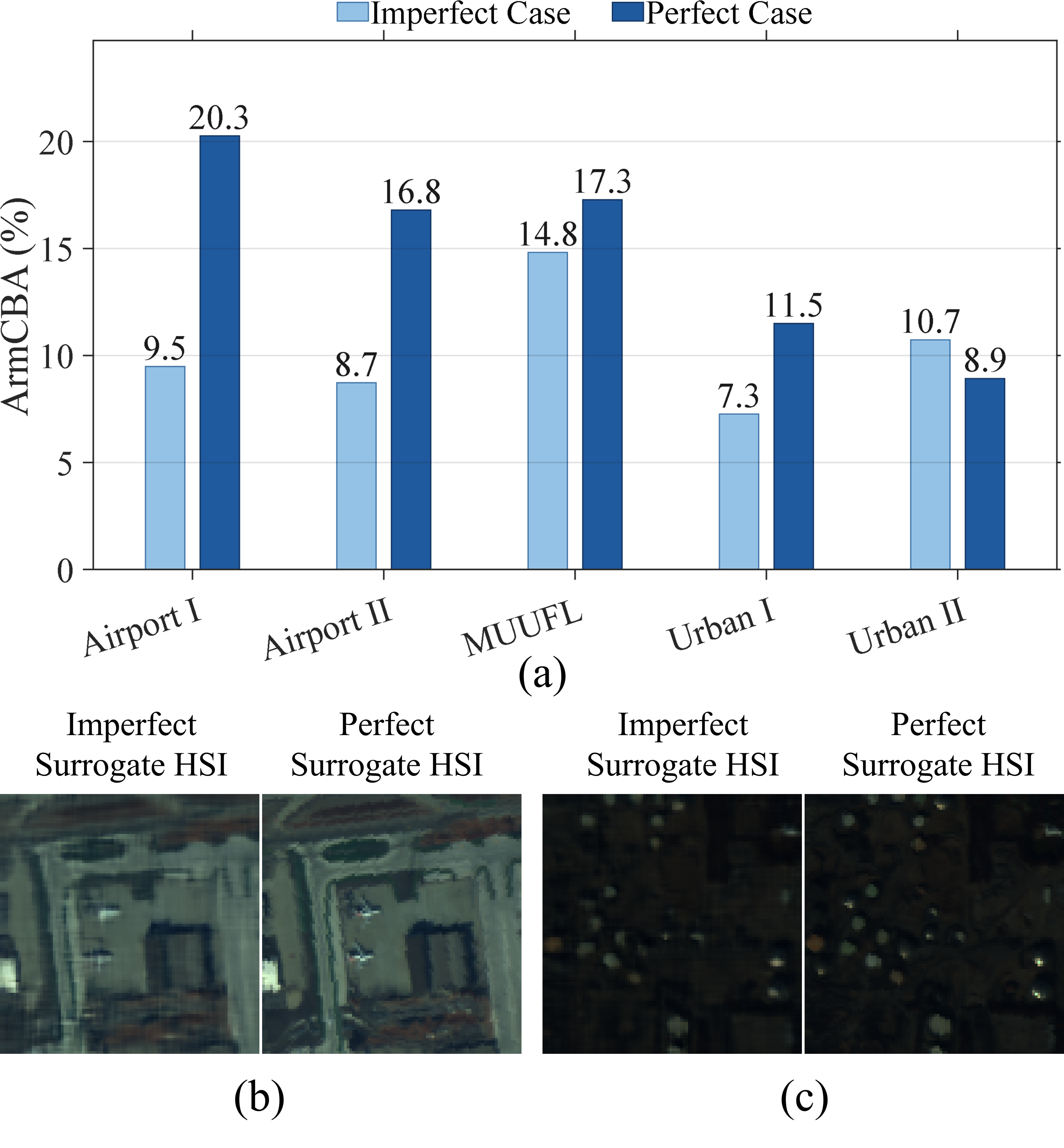}
    \caption{(a) Experimental comparisons of AHAD performance under perfect and imperfect surrogate-HSI conditions.
    (b) and (c) Qualitative comparisons between the perfect and imperfect surrogate HSIs on the Airport \uppercase\expandafter{\romannumeral 1} and Urban \uppercase\expandafter{\romannumeral 2} datasets, respectively. 
    The results demonstrate that the quality of surrogate HSIs affects the resulting AHAD performance.
    Further analyses are provided in Section \ref{subsubsec: Surrogate error}.
    }\label{fig: HDIP}
    \vspace{-0.3cm}
\end{figure}
%
To produce anti-detection signals for AHAD, a surrogate HSI $\widetilde{\mathbf{Y}}_H$ that approximates the inaccessible $\mathbf{Y}_H$ is generally essential.
Although $\widetilde{\mathbf{Y}}_H$ can be reasonably obtained using multiple contemporary techniques like HSI synthesis \cite{10000408} and spectral reconstruction \cite{young2026spectral} ({see} discussions in Section \ref{sec: Intro}), potential errors introduced at each reconstruction stage may accumulate to the next step and consequently propagate to the resulting surrogate HSIs, thereby rendering them insufficiently reliable in approximating $\mathbf{Y}_H$.     
Accordingly, investigating the influence of surrogate-HSI quality on this challenging AHAD task is insightful.
To this end, we conduct a simulation experiment to assess AHAD performance under perfect and imperfect surrogate-HSI conditions.
First, under perfect conditions, the original HAD datasets can be used straightforwardly as surrogate HSIs.
Second, for the imperfect condition, we adopt an iterative masking-and-reconstruction strategy to simulate potential error-accumulation effects.

Specifically, in practice, the reconstruction of surrogate HSI may largely rely on AI-based modules.
Accordingly, artifacts or residual errors introduced by a representative DL network can be regarded as a reasonable proxy for potential error arising from real-world surrogate-HSI reconstruction. 
In this experiment, we use the hyperspectral deep image prior (HDIP) \cite{HDIP} as the DL network, because this single-data, unsupervised method does not rely on additional training data, thereby preventing the assessment from being biased by the data quality or distribution.
Before this HDIP reconstruction step, 60\% of the elements of the input HSI are randomly masked to simulate limited input information encountered in a certain module.
For example, when conducting super-resolution, only low-resolution observations are available while the high-resolution details remain unavailable. 
Furthermore, since practical surrogate-HSI reconstruction may involve multiple sequential stages, we iteratively apply this HDIP reconstruction step (i.e., masking-and-reconstruction), with the output of the previous step as the input of the subsequent step.
Hence, reconstruction errors are accumulated and propagated to the output of the last HDIP step, which serves as the imperfect surrogate HSI.
In this experiment, the number of HDIP steps is set to 3.
Accordingly, the perfect and imperfect surrogate HSIs can be used to separately generate the corresponding AHAD signals.
 To better isolate the influence of surrogate-HSI quality, all evaluations are conducted under the zero-shifting case on the five real-world datasets.
The AHAD performance in terms of the ArmCBA score is averaged across the top-performing HAD baselines.
As demonstrated in {Fig.}~\ref{fig: HDIP}(a), imperfect surrogate HSIs generally degrade the ArmCBA scores, indicating the importance of their reconstruction fidelity.
Taking the Airport \uppercase\expandafter{\romannumeral 1} dataset as an example, the imperfect surrogate HSI reveals noticeable accumulated artifacts in both background and anomaly components, as illustrated in {Fig.}~\ref{fig: HDIP}(b).
Such pronounced artifacts may distort the overall data characteristics and disrupt the intrinsic relationship between background and anomaly signatures.
Accordingly, AHAD signals estimated from this less reliable surrogate HSI exhibit limited AHAD performance compared with those obtained from their perfect-condition counterparts.  
By contrast, a high-fidelity surrogate-HSI reconstruction can yield an anti-detection signal that achieves AHAD performance competitive with that of perfect conditions, as shown by results on the MUUFL and Urban \uppercase\expandafter{\romannumeral 2} datasets.
Among these cases, we note that the AHAD performance is slightly better under imperfect conditions on the Urban \uppercase\expandafter{\romannumeral 2} dataset, which may be attributed to the implicit prior introduced by HDIP.
However, we regard this as a data-specific result, rather than a general observation.

{\blue According to the above evaluations, the quality of surrogate HSIs does influence the resulting AHAD performance.
Reconstructions with higher fidelity generally yield AHAD performance closer to that achieved under the ideal surrogate-HSI condition.
Nevertheless, these findings should be interpreted in the context of current simulation settings.
As mentioned earlier, we use the random mask-and-reconstruction strategy to simulate incomplete input information. 
The introduced AI artifacts mainly illustrate error accumulation in the surrogate HSI reconstructions while keeping the CSI condition ideal.
In practical AHAD applications, jointly fine-tuning of the integrated AI modules can mitigate potential error accumulation, thereby enhancing the reliability of the resulting AHAD signals.
However, more realistic imperfect CSI conditions, such as spectral range differences and viewing-geometry changes, should be jointly considered in future work.
For instance, when the spectral range of target sensors is only approximately known, using a surrogate HSI with a broader wavelength range may yield more promising AHAD results, as it covers most of the spectral bands captured by the actual sensor.
}

\subsubsection{Stealthiness Analysis and Baseline Comparison}\label{subsubsec: p-bnemchmark}
\begin{figure}[t]
    \centering
    \includegraphics[width=1\linewidth]{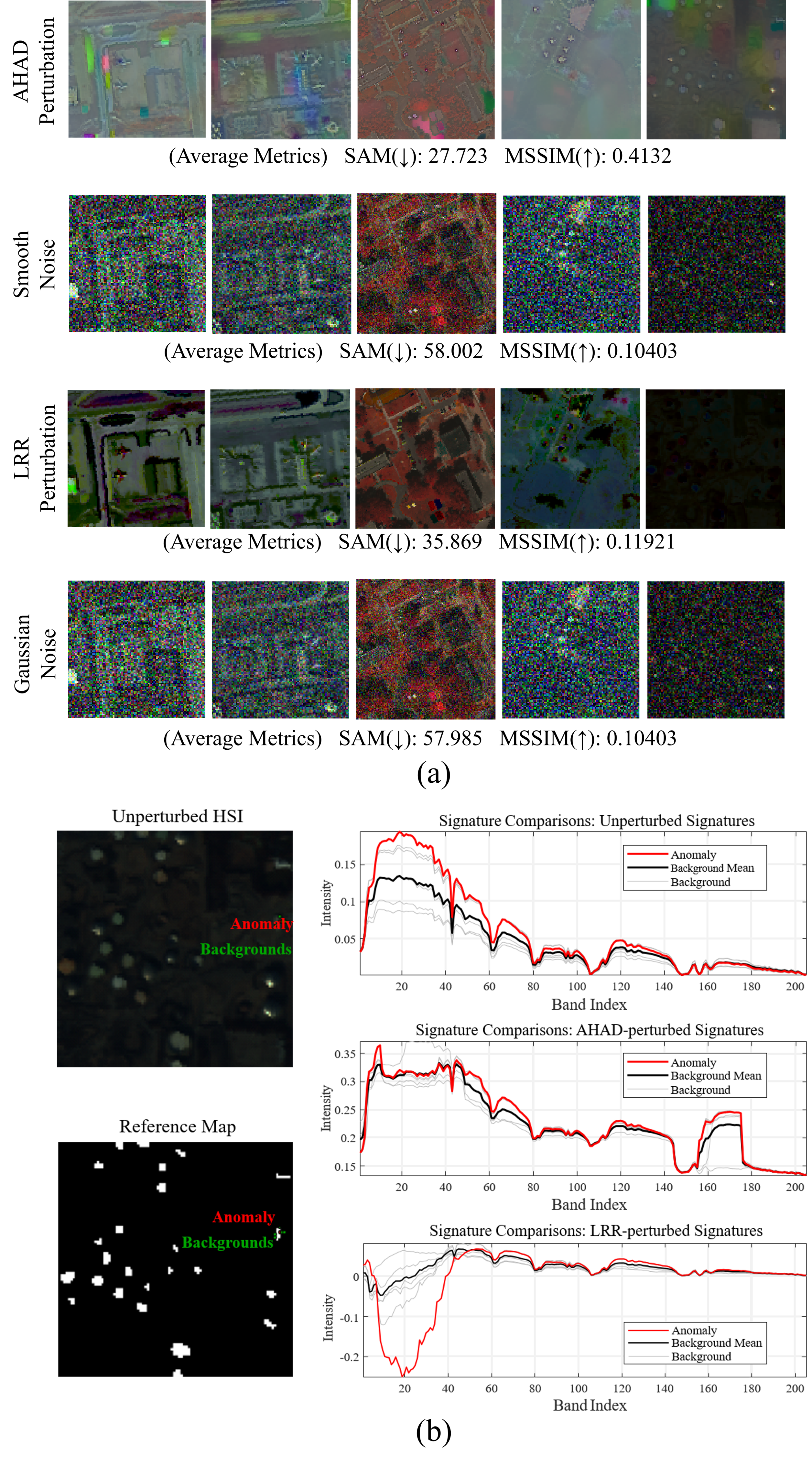}
    \caption{(a) Qualitative and quantitative comparisons of the HSIs perturbed by the proposed AHAD framework and the baseline strategies at comparable energy levels across the five real-world datasets.
    (b) Comparisons between the anomaly signature and the surrounding background signatures under unperturbed, AHAD-perturbed, and LRR-perturbed conditions, where the spatial locations of the extracted signatures are indicated in the HSI and reference maps.
    More detailed analyses are provided in Section \ref{subsubsec: p-bnemchmark}.
    }\label{fig: benchmark}
    \vspace{-0.3cm}
\end{figure}
\begin{figure}[t]
    \centering
    \includegraphics[width=1\linewidth]{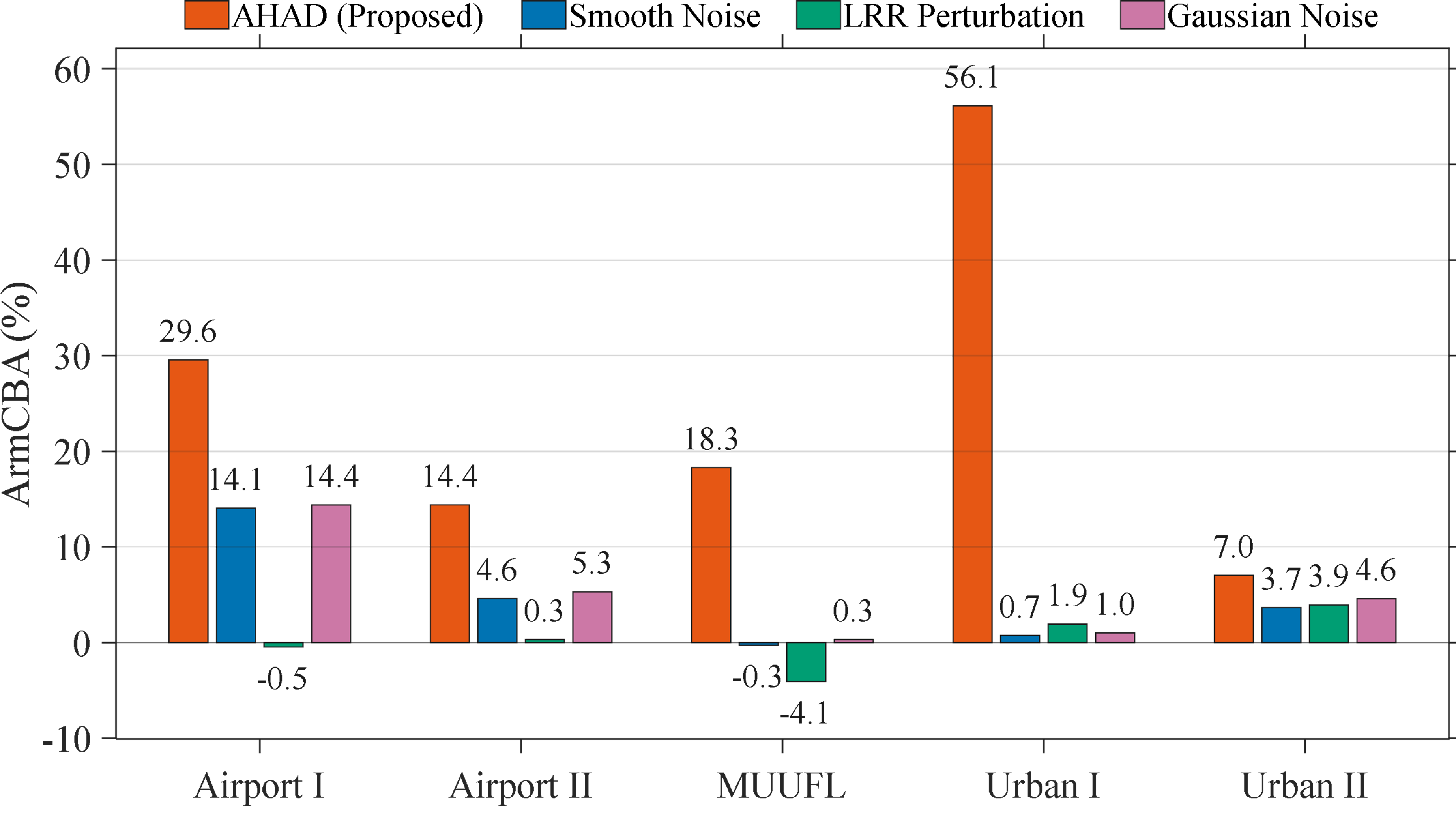}
    \caption{Quantitative comparison of the proposed AHAD framework with the baseline perturbation strategies illustrated in {Fig.}~\ref{fig: benchmark}(a), in terms of ArmCBA across the five real-world datasets.
    The proposed AHAD framework consistently achieves higher ArmCBA scores than the baseline perturbations.
    Detailed experimental settings and analyses are presented in Section \ref{subsubsec: p-bnemchmark}.
    }\label{fig: benchmark_AHAD}
    \vspace{-0.3cm}
\end{figure}
%
This section provides additional comparisons with baseline perturbations to demonstrate the effectiveness and stealthiness of the proposed framework.
Since this study presents the first AHAD paper, no existing AHAD algorithms or established methods are publicly available for direct comparisons. 
Accordingly, several commonly used perturbation strategies are adopted as benchmarks.
First, the representative Gaussian noise and its smoothed counterpart are incorporated as the noise-based baselines.
Specifically, a Gaussian noise tensor with the same dimensions as the target HSI is sampled from a standard normal distribution, after which a spectral-spatial Gaussian filter is applied.
Then, we subtract the global mean of the filtered noise tensor to enforce a zero-mean distribution, which prevents brightness and spectral offsets potentially caused by the filtering, thereby yielding smooth noise.
In this experiment, the Gaussian filter is implemented using the MATLAB function ``imgaussfilt3'' with a standard deviation of 0.25.

Furthermore, some meticulous readers may conjecture that the AHAD could be achieved by suppressing the anomaly component of the target HSI using LRR, as depicted in case (c) of {Fig.}~\ref{fig: intro_fig}.
To investigate this possibility, an LRR-based HAD algorithm is employed to decompose the anomaly tensor from the input HSI.
The anomaly tensor is first squared element-wise to emphasize anomalous components and then subtracted from the input HSI.
Nevertheless, a single-step anomaly suppression has only a negligible impact on this challenging AHAD task.
Hence, we fed the resulting anomaly-suppressed HSI into the next LRR iteration and again conducted the anomaly subtraction mentioned above.
This strategy is referred to as iterative LRR anomaly suppression.   
After all LRR anomaly suppressions are completed, the difference between the final anomaly-suppressed HSI and the original input HSI is defined as the LRR perturbation.
Here, LRASR \cite{LRASR} is adopted as the LRR-based HAD method, with the number of iterations set to 5.
For fair comparisons, all baseline perturbations, including Gaussian noise, smooth noise, and LRR perturbation, are normalized to have a comparable energy level to the proposed AHAD signal.

Initially, to demonstrate the stealthiness of the proposed AHAD framework, the qualitative and quantitative comparisons of HSIs perturbed by different perturbation strategies are shown in {Fig.}~\ref{fig: benchmark}(a).
As demonstrated, the noise-based perturbations introduce pronounced, unfavorable patterns, leading to unnatural appearance and limited stealthiness.   
In contrast, both the AHAD-perturbed and LRR-perturbed HSIs greatly preserve more realistic image characteristics of the original scenes.
However, their attributes remain distinguishable; the proposed AHAD framework introduces hazy-like patterns to achieve ARAB.
LRR perturbations generally concentrate the energy in local areas or certain targets.  
To further quantify the difference between these perturbation strategies, we employ two full-reference metrics, spectral angle mapper (SAM) \cite{HyperKING} and mean structural similarity index (MSSIM) \cite{CODE}, to compare them with their corresponding unperturbed HSIs. 
The SAM $(\downarrow)$ and MSSIM $(\uparrow)$ averaged across the five real-world datasets are shown below each perturbation strategy in {Fig.}~\ref{fig: benchmark}(a).
The results show that the proposed AHAD framework best preserves realistic spatial-spectral characteristics of the original HSIs, as substantiated by its SAM and MSSIM scores compared with all baseline perturbation strategies. 

Nevertheless, in real-world AHAD scenarios, unperturbed HSIs are generally unavailable for evaluating the stealthiness of different perturbation strategies.
Accordingly, a reliable AHAD technique should reduce the discrepancy between anomaly and background signatures (i.e., ARAB) without introducing unfavorable patterns (e.g., noise or impulse distortions), rather than preserving all characteristics of the original HSIs.
Clearly, the noise-based perturbation strategies fail to satisfy these requirements.
Therefore, we compare the anomaly and background signatures obtained by the LRR strategy and the proposed AHAD framework.
As shown by a representative example in {Fig.}~\ref{fig: benchmark}(b), the LRR perturbation strategy not only fails to assimilate the anomalous signatures into the background but also introduces a more pronounced discrepancy between anomaly and background signatures, which potentially compromises AHAD performance. 
In contrast, although the proposed AHAD framework may modify the original signatures, both resulting anomaly and background signatures maintain realistic spectral characteristics (e.g., spectral smoothness), while the anomaly signature is effectively assimilated into its surrounding background (i.e., ARAB).
To investigate the effectiveness of these perturbation strategies, we further conduct AHAD experiments against the top-performing HAD methods.
However, in practice reconnaissance scenarios, severely noisy HSIs, such as those shown in {Fig.}~\ref{fig: benchmark}(a), are unlikely to be used directly for HAD without appropriate restoration or denoising processing.
Therefore, to ensure practical yet fair comparisons, HSIs perturbed by each strategy are processed by the hyperspectral denoiser described in Section \ref{subsec: Resistance} before performing HAD.
The resulting AHAD performance is measured by ArmCBA and averaged across all top-performing HAD methods.
As observed in {Fig.}~\ref{fig: benchmark_AHAD}, the proposed AHAD framework consistently and substantially outperforms all baseline perturbation strategies. 
In addition, we observe that LRR perturbations can yield negative ArmCBA scores, indicating that $\text{AUC}_{ap}$ can even exceed $\text{AUC}_{up}$ after LRR perturbation. 
This observation echoes the preceding signature comparisons, suggesting that LRR perturbations may instead increase the discrepancy between anomaly and background signatures.
Overall, based on the above analyses, the proposed AHAD framework achieves the most favorable balance between perturbation stealthiness and AHAD effectiveness compared with commonly used perturbation strategies. 
\subsubsection{Robustness under Increasing CSI Uncertainty}\label{subsubsec: larger pixel-shift}
\begin{figure}[t]
    \centering
    \includegraphics[width=1\linewidth]{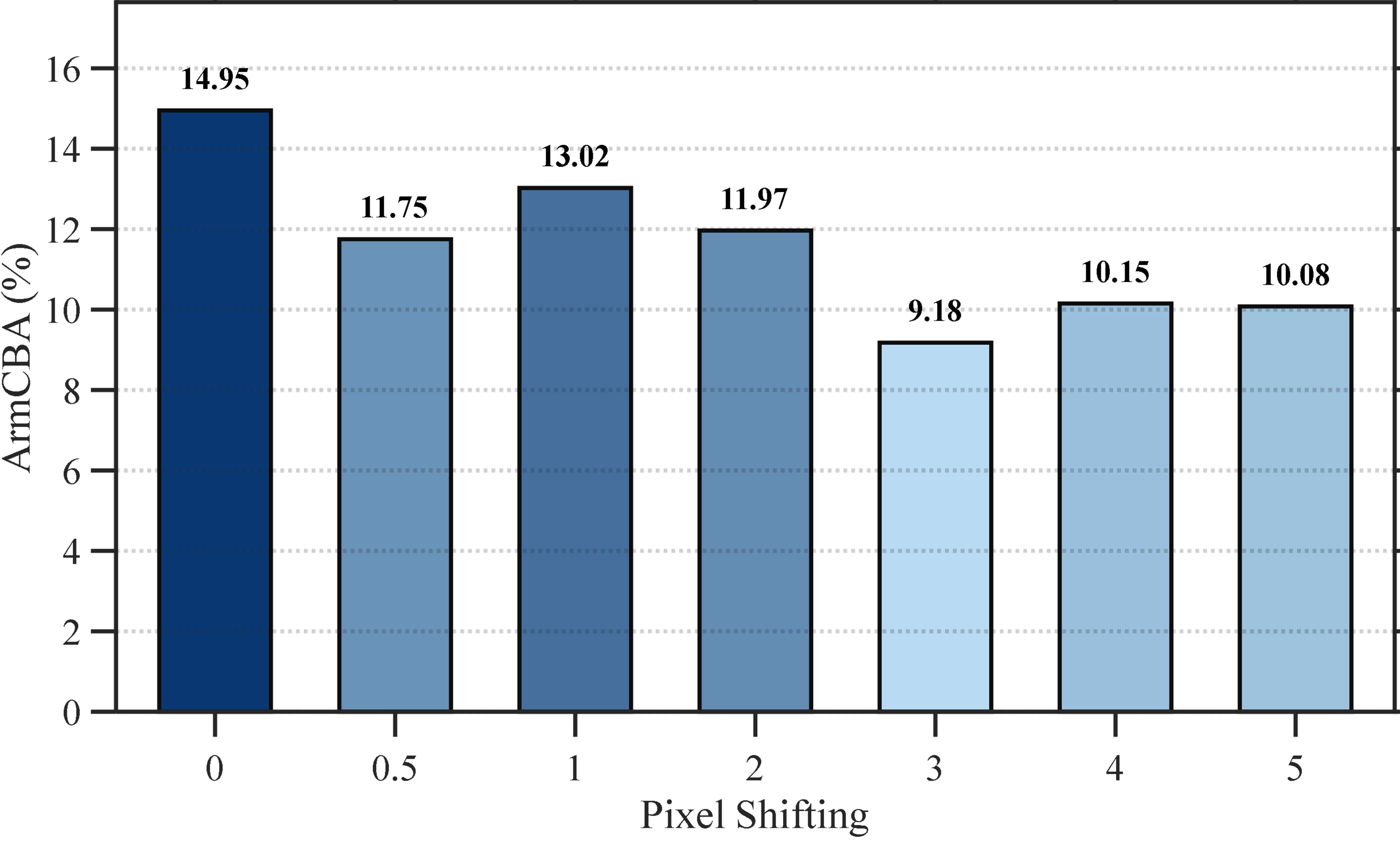}
    \caption{Robustness evaluation of the proposed AHAD framework under increasing pixel-level shifting and sub-pixel-level misalignment.
    As demonstrated, although larger pixel-level shifting generally degrades the average ArmCBA scores, the proposed AHAD framework exhibits considerable robustness, retaining over 60\% of its zero-shift performance.
    Moreover, the proposed AHAD framework demonstrates consistent robustness against sub-pixel misalignment.
    Further analyses and discussion are presented in Section \ref{subsubsec: larger pixel-shift}.
    }\label{fig: Larger_pixel_shifting}
    \vspace{-0.3cm}
\end{figure}
%
In this section, we further extend the evaluation to more challenging scenarios to investigate the robustness of the proposed AHAD framework.
First, we conduct evaluations under increasingly severe pixel-level shifts, with the number of shifted pixels $r$ varying from zero to five. 
For each pixel-shifting condition $r$, the set of candidate pixel-shifting contains $(2r+1)^{2}$ cases [{see} \eqref{eq: S}].
To impose progressively larger spatial misalignment, we set $\boldsymbol{\Delta}:=(r,r), ~\forall~ r=1,\cdots5$.
Experiments under these pixel-shifting conditions are conducted by following the settings described in the third paragraph of Section \ref{subsec: exp setting}.
For the zero- and one-pixel-shifting cases, we reuse the corresponding results reported in Section \ref{subsec: main exp} to avoid conducting redundant experiments.
To clearly characterize the effect of pixel-shifting on the proposed AHAD framework, the reported performances obtained at each pixel-shifting $r$ are averaged across the five real-world datasets and the top-performing HAD methods.

The results are presented in {Fig.}~\ref{fig: Larger_pixel_shifting}, wherein the integer and fractional shifting magnitudes correspond to pixel-level shifting and sub-pixel-level misalignment (to be detailed later), respectively.
For the progressively increasing pixel-level shifting, the corresponding ArmCBA scores achieve approximately 15\% under relatively small $r$, indicating promising AHAD performance ({see} Remark \ref{remark 2}).
As $r$ increases, each additional shifted pixel generally reduces the average ArmCBA score by around one to two points.  
Nevertheless, when the pixel-shifting increases from four to five pixels, the average ArmCBA score is retained at approximately 10\%, rather than further decreasing, revealing considerable robustness of the proposed AHAD framework.
Overall, the results under the pixel-level shifting in {Fig.}~\ref{fig: Larger_pixel_shifting} show that the proposed AHAD framework can retain over 60\% of its zero-shifting performance even under larger shifting magnitude. 

Subsequently, we evaluate the robustness of the proposed AHAD framework against sub-pixel-level misalignment.
To this end, we first introduce the sub-pixel misalignment effect into the surrogate HSI by extending the approach described in \cite{957286}.
In detail, the original surrogate HSI is first spatially upsampled by a factor of $M$ using bicubic interpolation, then shifted by $r$ pixels at the resulting resolution by the same steps described in Section \ref{subsec: exp setting}, and consequently downsampled back to its original spatial resolution.
Accordingly, a pixel-level shifting magnitude $\boldsymbol{\Delta}$ of $(r,r)$ in the upsampled HSI corresponds to a fractional pixel-shifting of $(\frac{r}{M}, \frac{r}{M})$ at the original resolution, thereby introducing the desired sub-pixel misalignment for robustness assessments.

Nevertheless, an excessively small sub-pixel displacement, such as 0.1 pixels, may have only a marginal effect in this evaluation, whereas an overly large displacement (e.g., 0.9) is close to pixel-level shifting cases that have been investigated above. 
Therefore, we set the sub-pixel misalignment to $\frac{r}{M}:=0.5$ pixels, with $M:=2$ and $r:=1$, in this experiment to enable a sufficiently representative and discriminative assessment. 
As demonstrated in {Fig.}~\ref{fig: Larger_pixel_shifting}, although the proposed AHAD framework is primarily designed to address pixel-level shiftings, it remains considerably robust against such sub-pixel misalignment. 
Specifically, the framework retains over 78\% of its zero-shifting performance in the ArmCBA score, which is comparable to that obtained under relatively small pixel-shifting conditions.
In summary, the above experimental results support the robustness of the proposed AHAD framework under both pixel-level shifting and sub-pixel-level misalignments.

\subsection{AHAD against Foundation Model-Based Detector}\label{subsubsec: HyperSIGMA}
\begin{figure}[t]
    \centering
    \includegraphics[width=1\linewidth]{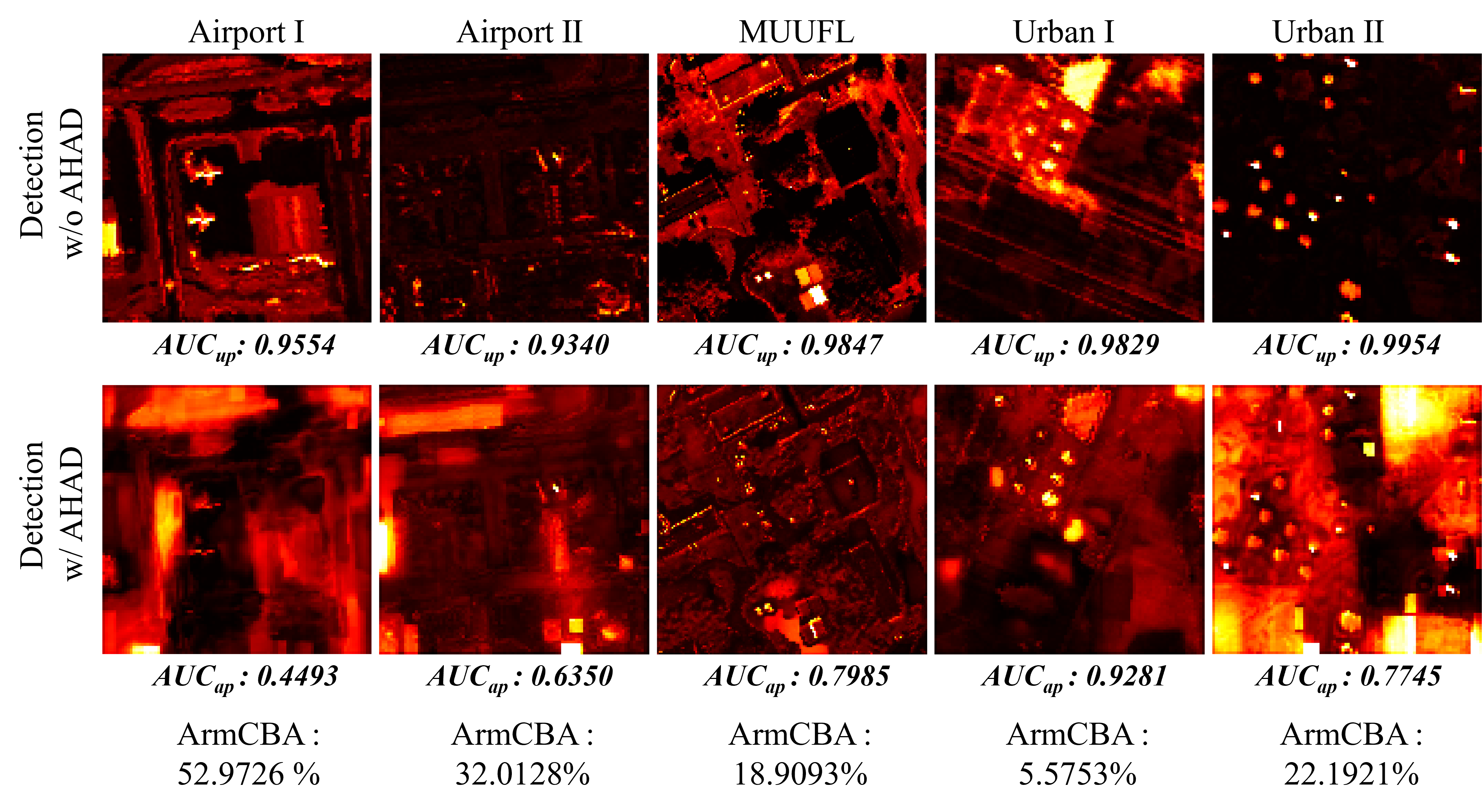}
    \caption{Experimental comparisons of the HyperSIGMA-based anomaly detector \cite{HyperSIGMA} without (w/o) and with (w/) the AHAD perturbation.
    As shown in the upper row, such a HyperSIGMA-based method achieves an average $\text{AUC}_{up}$ of 0.9705, outperforming the customized HAD approaches reported in Table \ref{tab: quantitative} and demonstrating the strong transferability of the hyperspectral foundation model to the HAD task.
    Nevertheless, the proposed AHAD technique remains effective against this advanced detector across most scenes, resulting in an average $\text{AUC}_{ap}$ of 0.7171 and an average ArmCBA score of 26.3324\%.
    More detailed analyses and discussion are presented in Section \ref{subsubsec: HyperSIGMA}. 
    }\label{fig: HyperSIGMA_HAD}
    \vspace{-0.3cm}
\end{figure}
With the rapid advancement of deep learning models and AI techniques, recent research attention has increasingly shifted toward building networks with generalizable capabilities rather than training them for specific tasks. 
Large-scale models that are pretrained on massive datasets and then transferred to a wide range of downstream tasks are commonly known as foundation models.
Representative examples include bidirectional encoder representations from transformers (BERT) \cite{BERT}, masked autoencoder (MAE) \cite{MAE}, and segment anything \cite{kirillov2023segment}.
Recently, foundation models have also been introduced to the RS community for supporting downstream applications, such as SpectralGPT \cite{SpectralGPT} and Skysense++ \cite{Skysens++}.
Consequently, beyond evaluating the proposed AHAD technique against mainstream customized HAD methods, as reported in Table \ref{tab: quantitative}, we further extend our experiment to foundation model-based HAD methods to provide a more thorough evaluation.

In the following experiment, a recent transformer-based hyperspectral foundation model, hyperspectral intelligence comprehension foundation
model (HyperSIGMA) \cite{HyperSIGMA}, is incorporated as the baseline. 
The backbone of HyperSIGMA was pretrained on the HyperGlobal-450K dataset \cite{HyperSIGMA}, which comprises 247,072 HSIs cropped from EO-1 imagery and 200,000 cropped from GF-5 imagery.
This pretraining paradigm enables HyperSIGMA to have strong transferability across a broad range of downstream hyperspectral tasks, including HAD \cite{HyperSIGMA}.
According to the official implementation, the pretrained spectral and spatial transformers in HyperSIGMA are jointly adopted as the hyperspectral feature encoder, followed by a $1\times 1$ 2D convolutional layer serving as the detection head.
This HAD configuration is referred to as the ``SSHADFramework'' in the official implementation.    
To transfer HyperSIGMA to the HAD task, cross-entropy loss is employed as the training objective, with Mahalanobis-distance-based anomaly scores \cite{RX} derived from the input HSI serving as the pseudo-labels.
The training epochs, learning rates of the AdamW optimizer, and weight decay are set to $1\times 10^{1}$, $6\times 10^{-5}$, $5\times 10^{-4}$, respectively, across the five real-world datasets ({see} {Fig.}~\ref{fig: dataset}).
The experimental settings of the proposed AHAD framework remain unchanged, as presented in Section \ref{subsec: exp setting}.
To account for CSI uncertainty when evaluating the framework against this foundation model-based detector, the pixel-shifting case is set to (1, 1) in this experiment.
With the above configurations, the HyperSIGMA-based detector achieves strong performance, yielding a higher average $\text{AUC}_{up}$ than the customized HAD methods reported in Table \ref{tab: quantitative}, as demonstrated in the upper row of {Fig.}~\ref{fig: HyperSIGMA_HAD}. 
Despite its superior detectability on unperturbed HSIs, the proposed AHAD technique remains effective against such a foundation model-based detector across most scenes.
For example, the anomalous targets, aircraft, become barely distinguishable in the resulting AHAD-perturbed detection results, as presented in the lower row of {Fig.}~\ref{fig: HyperSIGMA_HAD}. 
Although the AHAD effect may be relatively less pronounced on the Urban \uppercase\expandafter{\romannumeral 1} dataset, potentially due to the inherent data characteristics, the anti-detection signal can still achieve ArmCBA scores of approximately 20\% or higher, suggesting its effectiveness against this foundation model-based method.    
We attribute this observation to two potential factors: the pretraining strategy of HyperSIGMA and its post-adaptation for HAD.
Specifically, although the foundation model exhibits strong generalization and transferability, the pretraining was conducted mostly on high-quality data that do not account for such anti-detection perturbations.
Consequently, when serving as a hyperspectral feature encoder, the backbones of HyperSIGMA may have limited capability to suppress the AHAD perturbation. 
Second, due to the lack of ground-truth labels in the HAD task, adapting HyperSIGMA to the HAD task inherently relies on the pseudo-labels produced using conventional HAD methods.
However, mainstream HAD methods could be severely degraded by the AHAD perturbation signal ({see} Section \ref{subsec: main exp}), thereby compromising their reliability when serving as the supervision targets. 
In a nutshell, based on the experimental results, the proposed AHAD framework remains effective against most mainstream HAD methods as well as the HyperSIGMA-based detector.

\section{Conclusion and Future Works}\label{sec: Conclusion}
HSI has been recognized as a powerful RS information for detecting suspicious or anomalous objects, thereby facilitating the development of a broad range of HAD algorithms.
Nevertheless, such advanced techniques are a double-edged sword, fully exposing ground facilities within a target area.
To address this RS dilemma, this paper, for the first time, explicitly defines the concept of anti-detection against various HAD techniques ({see} {Fig.}~\ref{fig: AHAD_concept}).
We propose novel regularization mathematics to achieve this critical anti-detection mission, without requiring perfect CSI or prior knowledge of the HAD detector.
As a side contribution, a new quantitative performance index, ArmCBA, is proposed to evaluate the robustness of an HAD method against our AHAD signal.

Specifically, we formally define the AHAD problem characteristics (C1)-(C4), i.e., black-box uncertainty, ARAB property, EE, and RR ({see} Section \ref{subsec: perfect CSI Formulation}).
However, existing perturbation/interference strategies ({see} {Fig.}~\ref{fig: intro_fig}) violate at least one of the characteristics ({see} Section \ref{subsec: Implementation}), rendering them unreliable for addressing this security-critical problem.
To develop an unsupervised and robust AHAD method ({see} Lemma \ref{Lemma 1}), we perform anti-detection on the topology-enhanced HSI, rather than the original HSI, because successfully assimilating the anomalies w.r.t. the topology-enhanced case implies a highly effective ARAB procedure (namely, anomalies must also be assimilated w.r.t. the original (non-enhanced) HSI).
Comprehensive experiments show that the resulting robust anti-detection signal $\mathbf{P}^{\star}$ can degrade the HAD performance of a wide range of mainstream HAD algorithms ({see} Section \ref{subsec: main exp}) and the advanced foundation model-based method ({see} Section \ref{subsubsec: HyperSIGMA}), although its effectiveness may vary across the scenarios given the early-stage development of AHAD.
Quantitatively, the anti-detection signal induces a performance drop of approximately 16\% or even more in terms of average ArmCBA score ({see} Table \ref{tab: quantitative}), demonstrating strong potential for the future advancement of reliable AHAD that can prevent the detectors from collecting meaningful reconnaissance information.
Remarkably, although our anti-detection signal $\mathbf{P}^{\star}$ is not so weak as to be unobservable, experiments demonstrate that even when the detectors are aware of the anti-detection behavior and hence tries to remove the signal $\mathbf{P}^{\star}$ via some benchmark restoration technique, these mainstream HAD detectors could suffer from more serious performance drops ({see} Table \ref{tab: ablation}).
In addition, when compared with the commonly used baseline perturbation strategies, including noise and LRR anomaly suppression, the proposed AHAD framework consistently achieves the most favorable balance between perturbation stealthiness and AHAD effectiveness at a comparable energy level ({see} Section \ref{subsubsec: p-bnemchmark}).

This work demonstrates the technical feasibility of AHAD, inducing numerous important future research lines:
\begin{enumerate}
    \item 
    Currently, the anti-detection signals still require perceptible energy, so designing advanced regularization schemes to further facilitate EE is a problem worth investigating.
    Since DL networks themselves can be regarded as implicit regularization, leading to the seminal concept of deep image prior (DIP).
    Accordingly, advanced DL techniques, such as QUEENs \cite{HyperKING} and quantum DIP \cite{GQmu,PRIME}, may be used to regularize AHAD estimation and suppress energy redundancy.
    We call this future line unsupervised DL-based AHAD solutions.
    
    \item 
    Conversely, training an AHAD network using supervised learning (SL) can be envisioned as an acceleration strategy, namely SL-based AHAD.
    Specifically, the proposed AHAD framework enables offline generation of the high-quality HSI-perturbation pairs ($\mathbf{Y}_H$, $\mathbf{P}^{\star}$) using publicly available HSI datasets \cite{young2026spectral}.
    With these supervision targets (i.e., HSI-perturbation pairs), SL enables a neural network $f_{\theta}$ to approximate the mapping and predict the anti-detection signal through a single feed-forward pass [i.e., $\mathbf{P}^{\star}\approx f_{\theta}(\mathbf{Y}_H)$], instead of performing time-consuming iterative optimization for each dataset.
    Such a strategy may also alleviate the current limitation in parameter tuning, as the adjustment has been completed during the generation of training data.
    SL may further impose underlying advantages for such important AHAD tasks.
    For example, a large-scale pretrained model has shown potential to improve transfer performance on downstream tasks \cite{HyperSIGMA,MAEPrePretraining}.
    Accordingly, a supervised fine-tuned AHAD network may provide additional generalization or performance improvement compared to the current framework.
    These potential benefits remain to be empirically investigated.
    
    \item 
    Robust AHAD criteria may be further explored. 
    Although imperfect CSI has been considered to yield a robust solution, it preliminarily focuses on the single-detector scenario at this stage.
    Similar to multi-input multi-output (MIMO) systems in wireless communications, multiple detectors may simultaneously traverse the target area while sharing a unified anti-detection signal.
    In this context, multiple viewing geometries and different spatial misalignments must be considered simultaneously.
   For instance, a straightforward approach is to assign each detector a sensor-specific surrogate HSI and the corresponding misalignment condition.
    Accordingly, the current robust criterion can be extended by jointly optimizing the weighted sum of sensor-specific AHAD objective functions, thereby producing a unified AHAD signal for the more challenging multi-detector scenario.
    Nevertheless, different detector conditions (e.g., surrogate HSI and CSI uncertainty) may lead to conflicts among sensor-specific criteria, hindering the overall optimization process and preventing the resulting AHAD signal from achieving robust AHAD performance.
    To mitigate these challenges, a spatial mask or weighting strategy may be further introduced to explicitly concentrate the energy of anti-detection signals and pseudo-anomaly generation within regions that are important for AHAD performance across the surrogate HSIs, rather than enforcing global signal coverage over the entire target area.
    This strategy is expected to improve overall EE and mitigate potential data conflicts among sensor-specific surrogate HSIs.
    
    \item 
    Importantly, we observe that existing HAD methods are mainly robust against noise/outlier effects.
    Based on our proposed ArmCBA index, future HAD research can consider robustness against anti-detection perturbations.
    This will become essential, echoing the competing relationship between AA and adversarial defense. 
\end{enumerate}

\appendix

\subsection{Proof of Lemma \ref{Lemma 1}}\label{sec:proof Lemma 1}
In wireless communications, stochastic optimization strategies (e.g., minimizing the objective expectation) have been widely adopted to address uncertainty in robust beamforming \cite{chi2017convex}.
This fact motivates the formulation of a robust AHAD criterion in a probabilistic manner to address the CSI uncertainty.
To this end, we introduce a random variable $\boldsymbol{\Delta}_{\mathrm{R}}$ with a distribution $p$ to statistically characterize such uncertainty of pixel-shifting.
%
%
This leads to the following robust AHAD criterion, i.e.,
\begin{align}\label{eq: robust AHAD exp}
  \min_{\mathbf{P}}\quad 
& \mathbb{E}_{\boldsymbol{\Delta}_{\mathrm{R}}\sim p}\left[\mathcal{L}(\mathbf{P}|\mathcal{T}_{\boldsymbol{\Delta}_{\mathrm{R}}}(\widetilde{\mathbf{Y}}_H))\right]\notag
\\
\mathrm{s.t.}\quad 
& \widetilde{\mathbf{Y}}_A^{(\boldsymbol{\Delta}_{\mathrm{R}})}
\in [0,1]^{H\times W\times C}.
\end{align}

However, the probabilistic objective function and constraint hinder a straightforward optimization of \eqref{eq: robust AHAD exp}.
Hence, the Monte Carlo estimator \cite{MDL} is adopted to approximate the probabilistic expectation by a deterministic objective, i.e., 
\begin{align}\label{eq: MC}
\mathbb{E}_{\boldsymbol{\Delta}_{\mathrm{R}}\sim p}\left[\mathcal{L}(\mathbf{P}|\mathcal{T}_{\boldsymbol{\Delta}_{\mathrm{R}}}(\widetilde{\mathbf{Y}}_H))\right]
\approx
\frac{1}{m}\sum_{i=1}^{m} \mathcal{L}(\mathbf{P}|\mathcal{T}_{\boldsymbol{\Delta}^{(i)}}(\widetilde{\mathbf{Y}}_H)),
\end{align}
where $\boldsymbol{\Delta}^{(i)},\cdots,\boldsymbol{\Delta}^{(m)}$ denote independent and identically distributed (i.i.d.) shifting cases sampled by $p$.

Nevertheless, the unknown distribution $p$ fails the direct sampling steps using the true distribution.
Notably, the considered spatial misalignment is associated with pixel-level shifting, rather than uncountable sub-pixel-level misregistration.
This stems from the fact that the shifting arising from imperfect CSI is likely to be closer to cross-sensor misalignments, which are commonly modeled at the pixel level in the RS field \cite{11435906,10543175}. 
This property leads $p$ to a finite support $\mathcal{S}$ [{see} \eqref{eq: S}], allowing an explicit enumeration of all candidate samples.

Specifically, under $m$ trials, the number of repeated samples across each candidate is the frequency of its occurrences.
Accordingly, instead of performing a large number of trials via $p$, each shifting case in $\mathcal{S}$ can be evaluated once with some non-negative integer $\theta^{\boldsymbol{\Delta}}\geq0,~ \sum_{\boldsymbol{\Delta}\in\mathcal{S}} \theta^{\boldsymbol{\Delta}}=m$, i.e.,
\begin{align}\label{eq: MC2}
\frac{1}{m}\sum_{i=1}^{m} \mathcal{L}(\mathbf{P}|\mathcal{T}_{\boldsymbol{\Delta}^{(i)}}(\widetilde{\mathbf{Y}}_H))
=
\frac{1}{m}\sum_{\boldsymbol{\Delta}\in\mathcal{S}} \theta^{\boldsymbol{\Delta}} \mathcal{L}(\mathbf{P}|\mathcal{T}_{\boldsymbol{\Delta}}(\widetilde{\mathbf{Y}}_H)).
\end{align}
Let $q^{\boldsymbol{\Delta}}=\frac{1}{m}\theta^{\boldsymbol{\Delta}}\geq0,~ \sum_{\boldsymbol{\Delta}\in\mathcal{S}} q^{\boldsymbol{\Delta}}=1$ denote the normalized contributions with respect to $m$.
Moreover, the probabilistic box constraint in \eqref{eq: robust AHAD exp} can be transformed into a deterministic one by considering all candidate shifting cases.
Consequently, by ignoring irrelevant constants, the robust AHAD criterion can be expressed as 
\begin{align}
  \min_{\mathbf{P}}\quad 
& \sum_{\boldsymbol{\Delta}\in\mathcal{S}}
q^{(\boldsymbol{\Delta})}
\mathcal{L}\!\left(
\mathbf{P}
\mid
\mathcal{T}_{\boldsymbol{\Delta}}(\widetilde{\mathbf{Y}}_H)
\right)\notag
\\
\mathrm{s.t.}\quad 
& \widetilde{\mathbf{Y}}_A^{(\boldsymbol{\Delta})}
\in [0,1]^{H\times W\times C},\quad\forall\boldsymbol{\Delta}\in\mathcal{S}.
\end{align}
Therefore, the proof of Lemma \ref{Lemma 1} has been completed.
\hfill$\blacksquare$

\renewcommand{\thesubsection}{\Alph{subsection}}
\bibliographystyle{IEEEtran}
\bibliography{ref}

\begin{IEEEbiography}[{\resizebox{0.9in}{!}{\includegraphics[width=1in,height=1.25in,clip,keepaspectratio]{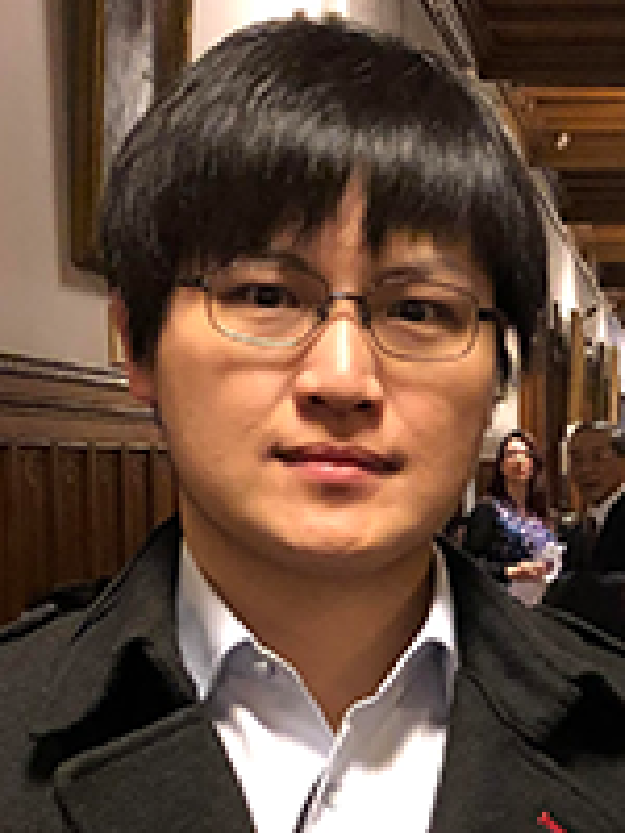}}}]
	{\bf Chia-Hsiang Lin}
(S'10-M'18-SM'24)
received the B.S. degree in electrical engineering and the Ph.D. degree in communications engineering from National Tsing Hua University (NTHU), Taiwan, in 2010 and 2016, respectively.
From 2015 to 2016, he was a Visiting Student of Virginia Tech,
Arlington, VA, USA.

He is currently a Professor with the Department of Electrical Engineering,
National Cheng Kung University (NCKU), Taiwan, and also serves as a Technical Director of Smart Sensing \& Systems Technology Center, Industrial Technology Research Institute (ITRI).
Before joining NCKU, he held research positions with The Chinese University of Hong Kong, HK (2014 and 2017),
NTHU (2016-2017),
and the University of Lisbon (ULisboa), Lisbon, Portugal (2017-2018).
He was an Assistant Professor with the Center for Space and Remote Sensing Research, National Central University, Taiwan, in 2018, a Visiting Professor with ULisboa, in 2019, and a Visiting Professor with Texas A\&M University, USA, in 2025 and in 2026.
His research interests include network science,
quantum computing,
convex geometry and optimization, blind signal processing, and imaging science.

Dr. Lin received the Emerging Young Scholar Award (The 2030 Cross-Generation Program) from National Science and Technology Council (NSTC), from 2023 to 2027,
the Outstanding Technical Achievement Award from IEEE Tainan Section, in 2026,
the Stanford World's Top 2\% Scientists, in 2026,
the Future Technology Award from NSTC, in 2022 and 2026,
the Outstanding Youth Electrical Engineer Award from The Chinese Institute of Electrical Engineering (CIEE), in 2022,
the Best Young Professional Member Award from IEEE Tainan Section, in 2021,
and the Prize Paper Award from IEEE Geoscience and Remote Sensing Society (GRS-S), in 2020.
He received the Ministry of Science and Technology (MOST) Young Scholar Fellowship, together with the EINSTEIN Grant Award, from 2018 to 2023.
In 2016, he was a recipient of the Outstanding Doctoral Dissertation Award from the Chinese Image Processing and Pattern Recognition Society and the Best Doctoral Dissertation Award from the IEEE GRS-S.
\end{IEEEbiography}

\begin{IEEEbiography}[{\resizebox{0.9in}{!}{\includegraphics[width=1in,height=1.25in,clip,keepaspectratio]{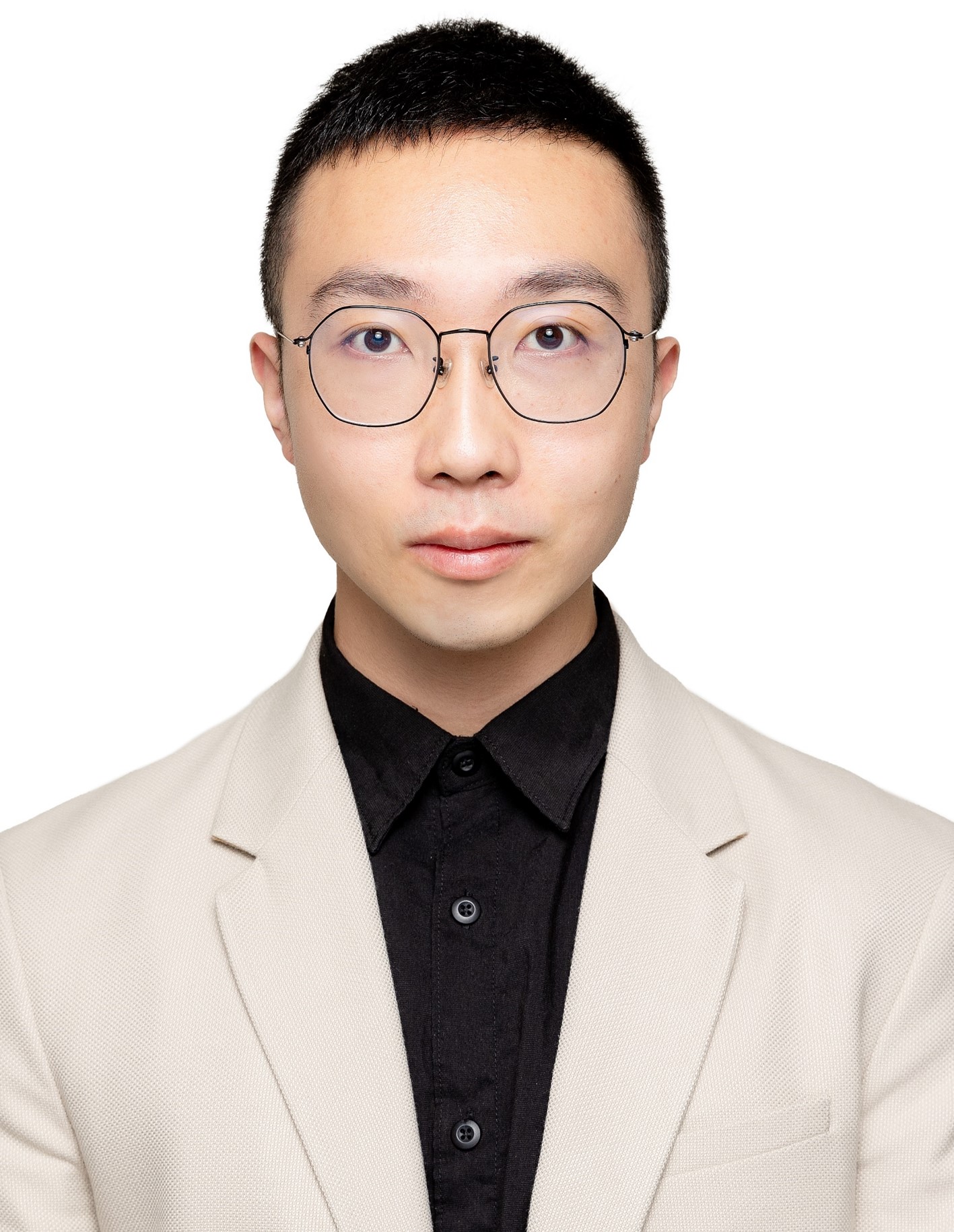}}}]
	{\bf Si-Sheng Young}
(S'23) is currently a Ph.D. student with the Intelligent
Hyperspectral Computing Laboratory (IHCL), Department of Electrical Engineering, National Cheng Kung University (NCKU), Tainan, Taiwan. 

In 2023, he received the Merit Award from The Grand Challenge ``Computing for the Future", Miin Wu School of Computing, NCKU, as well as the highly competitive ``Pan Wen Yuan Scholarship" from the Industrial Technology Research Institute (ITRI), Hsinchu, Taiwan.
In 2024, he received a highly competitive ``Scholarship Pilot Program to Cultivate Outstanding Doctoral Students" from the National Science and Technology Council (NSTC), Taiwan.
His research interests include convex optimization, deep learning, anomaly detection, data fusion, and imaging inverse problems.
\end{IEEEbiography}

\begin{IEEEbiography}[{\resizebox{0.9in}{!}{\includegraphics[width=1in,height=1.25in,clip,keepaspectratio]{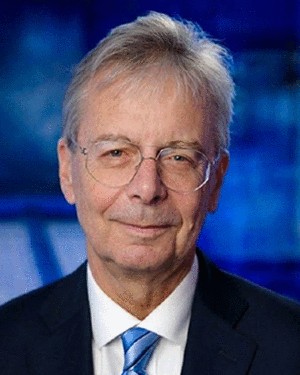}}}]
	{\bf Jon Atli Benediktsson} (Life Fellow, IEEE) received the Cand.Sci. degree in electrical engineering from the University of Iceland, Reykjavík, Iceland, in 1984, and the M.S.E.E. and Ph.D. degrees in electrical engineering from Purdue University, West Lafayette, IN, USA, in 1987 and 1990, respectively.

Since 1991, he has been with the University of Iceland and has been a Professor of electrical and computer engineering since 1996. 
He was the President and a Rector of the University of Iceland from 2015 to 2025 and the President of Aurora Universities from 2020 to 2024. 
His research interests are in remote sensing, biomedical analysis of signals, pattern recognition, image processing, and signal processing, and he has published extensively in those fields.

Dr. Benediktsson is an International Member of the U.S. National Academy of Engineering, a Fellow of SPIE, a member of the Academy of Europe, SigmaXi, and Tau Beta Pi. He was granted the IEEE Third Millennium Medal.
\end{IEEEbiography}

\end{document}